\documentclass[11pt]{article}
\usepackage{graphics,cite,amssymb,epsfig,float}
\usepackage[usenames,dvips]{color}
\usepackage{amsmath}
\usepackage{subfigure}
\usepackage{url}

\usepackage{multirow}
\makeatletter

\@addtoreset{equation}{section}
\makeatother

\def\lsim{\;\raise0.3ex\hbox{$<$\kern-0.75em\raise-1.1ex\hbox{$\sim$}}\;}
\def\gsim{\;\raise0.3ex\hbox{$>$\kern-0.75em\raise-1.1ex\hbox{$\sim$}}\;}

\def\ben{\begin{enumerate}}  \def\een{\end{enumerate}}
\def\bit{\begin{itemize}}    \def\eit{\end{itemize}}
\def\beq{\begin{equation}}   \def\eeq{\end{equation}}
\def\ba{\begin{array}}       \def\ea{\end{array}}
\def\bea{\begin{eqnarray}}   \def\eea{\end{eqnarray}}

\newcommand{\comment}[1]{}

\newcommand{\Br}{\text{Br}}

\begin{document}

\begin{titlepage}
\renewcommand{\thefootnote}{\fnsymbol{footnote}}
\setcounter{footnote}{0}

\vspace*{-2cm}
\begin{flushright}
LPT Orsay 12-36 \\

\vspace*{2mm}
\today
\end{flushright}

\vspace*{2mm}

\begin{center}
\vspace*{15mm}
{\Large\bf Enhancing lepton flavour violation in the supersymmetric inverse seesaw
beyond the dipole contribution} \\
\vspace{1cm}
{\bf Asmaa Abada\footnote{asmaa.abada@th.u-psud.fr}, Debottam Das\footnote{debottam.das@th.u-psud.fr}, Avelino Vicente\footnote{avelino.vicente@th.u-psud.fr} and C\'{e}dric Weiland\footnote{cedric.weiland@th.u-psud.fr}}

 \vspace*{.5cm} 
 Laboratoire de Physique Th\'eorique, CNRS -- UMR 8627, \\
Universit\'e Paris-Sud 11, F-91405 Orsay Cedex, France
\end{center}

\vspace*{10mm}
\begin{abstract}

\vspace*{3mm} In minimal supersymmetric models the $Z$-penguin usually
provides sub-dominant contributions to charged lepton flavour violating
observables.  In this study, we consider the supersymmetric inverse
seesaw in which the non-minimal particle content allows for dominant
contributions of the $Z$-penguin to several lepton flavour violating
observables.  In particular, and due to the low-scale (TeV) seesaw,
the penguin contribution to, for instance, $\Br(\mu \to 3e)$ and
$\mu-e$ conversion in nuclei, allows to render some of these
observables within future sensitivity reach. Moreover, we show that in
this framework, the $Z$-penguin exhibits the same non-decoupling behaviour which
had previously been identified in flavour violating Higgs decays
in the Minimal Supersymmetric Standard Model.

\end{abstract}

\vspace*{3mm}

\renewcommand{\thefootnote}{\arabic{footnote}}
\setcounter{footnote}{0}

\vspace*{5mm}
\end{titlepage}\setcounter{page}{2}

\section{Introduction}
\label{sec:intro}

In recent years, lepton physics has experienced an unprecedented
experimental development. A non-vanishing - even unexpectedly large -
value of the Chooz angle ($\theta_{13} \simeq 9 ^\circ$) has 
recently been measured by several (independent) collaborations 
\cite{Abe:2011fz,Hartz:2012pr,Adamson:2012rm,An:2012eh,Ahn:2012nd}. Such
a value of $\theta_{13}$ opens the door to very appealing
phenomenological possibilities, among which $CP$ violation in the
leptonic sector stands as the best example. In parallel to these
achievements, the other neutrino oscillation parameters (solar and
atmospheric) are being determined with very good precision
\cite{Tortola:2012te,Fogli:2012ua}. In the near future, one does
expect to identify the fundamental ingredients of flavour violation in
the neutral lepton sector, but on the other hand, flavour violation in the
charged lepton sector still remains to be observed. The present and
future generations of high-intensity facilities dedicated to
discovering flavour violation in charged lepton processes render
feasible the observation of such an event in the near future.

Despite the fact that minimal extensions of the Standard Model (SM)
can easily accommodate lepton flavour violation in the neutral lepton
sector (i.e. neutrino oscillations), the contributions of these models
to charged lepton flavour violating (cLFV) observables are typically
extremely small. On the other hand, when such models - for example,
the seesaw in its different realisations - are embedded within a
larger framework, one can expect large contributions to cLFV
observables, well within experimental reach.  This is the case of
supersymmetric versions of the seesaw mechanism, which in addition to
apport solutions to many theoretical and phenomenological issues, such
as the hierarchy problem, gauge coupling unification and dark matter,
can also account for neutrino data.

However, these scenarios have several caveats, the most upsetting one
being that they prove to be extremely hard to test, and thus can be
neither confirmed nor excluded. This stems from the fact that in order
to have sufficiently large Yukawa couplings (as required to account
for large cLFV branching ratios), the typical scale of the extra
particles (such as right-handed neutrinos, scalar or fermionic isospin
triplets) is in general very high, potentially close to the gauge
coupling unification scale.

This can be avoided if one simultaneously succeeds in having TeV-scale mediators, while preserving the possibility of large Yukawa couplings. 
From an effective theory point of view this is equivalent to the decoupling of the coefficients associated to the dimension-five (at the origin of neutrino masses) and dimension-six operators (for instance, four-fermion operators): in other words, decoupling the smallness of the light neutrino masses from the flavour violation sources. For instance, this is possible in the case of the type-II seesaw (and its supersymmetric (SUSY) realisations), as well as in the case of the so-called "inverse seesaw" (and the SUSY inverse seesaw).

The inverse seesaw \cite{Mohapatra:1986bd} constitutes a very
appealing alternative to the "standard" seesaw realisations, and has
recently been the subject of several dedicated
studies~\cite{Deppisch:2004fa,Deppisch:2005zm,Garayoa:2006xs,Arina:2008bb,Ma:2009gu,Dev:2009aw,Malinsky:2009df,Bazzocchi:2009kc,Hirsch:2009ra,Bazzocchi:2010dt,Catano:2012kw,Dias:2012xp,Hirsch:2012kv}. The
inverse seesaw can be embedded in the Minimal Supersymmetric
extension of the SM (MSSM) by the addition of two extra gauge singlet
superfields, with opposite lepton numbers ($+1$ and $-1$).  As
extensively discussed
in~\cite{Deppisch:2004fa,Deppisch:2005zm,Garayoa:2006xs,Arina:2008bb,Ma:2009gu,Dev:2009aw,Malinsky:2009df,Bazzocchi:2009kc,Hirsch:2009ra,Bazzocchi:2010dt,Catano:2012kw,Dias:2012xp,Hirsch:2012kv},
in this framework one can in principle have large neutrino Yukawa
couplings ($Y_\nu \sim {\mathcal{O}}(1)$) compatible with a seesaw
scale, $M_\text{seesaw}$, close to the electroweak one (and thus
within LHC reach), implying that there will be a significant
enhancement of cLFV observables.  This has fuelled a number of studies
focusing on the potentialities of the inverse seesaw regarding cLFV
and other phenomenological
issues~\cite{Deppisch:2004fa,Deppisch:2005zm,Garayoa:2006xs,Arina:2008bb,Dev:2009aw,Malinsky:2009df,Bazzocchi:2009kc,Hirsch:2009ra,Hirsch:2012kv,Mondal:2012jv,Dev:2012zg,Das:2012ze}.
As recently discussed~\cite{Abada:2011hm}, the contributions of the
comparatively light right-handed sneutrinos can enhance the
Higgs-mediated penguin diagrams, leading to an augmentation of some
observables - for instance $\Br(\tau \to 3 \mu)$ - by as much as two
orders of magnitude, and have a non-negligible impact on
Higgs-mediated leptonic B-meson decays and Higgs flavour violating
decays.

These unique features - when compared to other SUSY seesaw
realisations - open the door to rich phenomenological signatures, that
can be potentially tested in the near future.  The different high- and
low-energy phenomenological implications of this class of models
constitute the starting point to unveil the underlying mechanism of
lepton mixing.

There are currently a large number of
facilities~\cite{Aubert:2005wa,Aubert:2005ye,Bellgardt:1987du,Aubert:2003pc,Akeroyd:2004mj,Kuno:2005mm,Kiselev:2009zz,Ritt:2006cg,Hayasaka:2007vc},
dedicated to the search of processes such as rare radiative decays,
3-body decays and muon-electron conversion in nuclei. Likewise,
rare leptonic and semi-leptonic meson decays also offer a rich testing
ground to experimentally probe cLFV.

These (low-energy) searches are complementary to the LHC which, in
addition to directly searching for new physics states, also allow to
study numerous signals of cLFV at high-energy, typically in
association with neutralino-slepton decay chains.  In order to
disentangle the underlying model of lepton flavour violation, one
relies on numerous strategies based on the interplay of low- and
high-energy cLFV observables (see for
example~\cite{Abada:2010kj,Esteves:2010si,Abada:2011mg,Abada:2012re}).
However, there are other avenues that can be explored in this quest to
disentangle the underlying mechanism of neutrino mass generation, at
the origin of lepton flavour violation: this approach is based upon
exploring the correlation (or lack thereof) between different,
unrelated, low-energy cLFV observables. The distinctive features of
the underlying model will be manifest in the nature and specific
hierarchy of the different contributions. For instance, in SUSY models
where $\gamma$-penguins provide the dominant contribution to radiative
and 3-body cLFV decays, one expects a strict correlation between
$\Br(\mu \to e \gamma)$ and CR($\mu - e$, N). This is the case of
constrained Minimal Supersymmetry Standard Model (cMSSM) based
scenarios where additional lepton flavour violating sources have been
introduced. Deviations from strict universality (as is the case of
non-universal Higgs masses, NUHM), where for example Higgs-mediated
penguins might play a significant r\^ole in $\mu-e$ conversion, break
this strict correlation\cite{Arganda:2007jw}.

Independently of the specific mechanism of SUSY breaking, specific contributions to cLFV observables manifest a peculiar behaviour. A very interesting case is that of the $Z$-penguin: in the MSSM, the contributions of the $Z$-penguin to cLFV observables (such as $\ell_i \to 3\ell_j$ and $\mu-e$ conversion in nuclei) are suppressed by a subtle cancellation between the different terms in the amplitude.
However, it has recently been noticed~\cite{Hirsch:2012ax} that in models where new couplings are present or where the particle content is larger than that of the MSSM, the cancellation no longer holds, and the $Z$-penguin contributions can in fact provide the dominant contributions to cLFV processes such as 3-body decays, $\ell_i \to 3\ell_j$ and $\mu - e$ conversion in heavy nuclei. Models with additional couplings (as is the case of trilinear R-parity violating supersymmetric models) were discussed in~\cite{Hirsch:2012ax,Dreiner:2012mx}.

Contrary to the "standard SUSY seesaw", in the inverse SUSY seesaw the new states (in addition to those of the MSSM) do not decouple: indeed, the sterile states can be as light as to lie in the sub-GeV scale. In the framework of the inverse SUSY seesaw, one can see that the $Z$-penguins will also provide sizeable, if not dominant, contributions to a number of cLFV observables. 

In this work, we consider a realisation of the inverse seesaw, embedding it into an otherwise lepton flavour conserving supersymmetric extension of the SM, the cMSSM. We conduct a detailed study of the impact that enhanced $Z$-penguins might have on a large number of low-energy cLFV observables, in particular $\mu-e$ conversion and $\mu\to 3e$ decay which in addition to being greatly enhanced by the $Z$-penguin, also have the best experimental prospects concerning the expected future sensitivities. 
Our results reveal that for vast regions of the parameter space,  many cLFV observables are indeed boosted by the unsuppressed $Z$-penguin contribution and are within reach of present and future experiments.

Moreover, our analysis reveals that, similarly to what occurs in the MSSM,  flavour changing Higgs boson decays  (where the Higgs boson contributions do not decouple with increasing supersymmetric masses~\cite{Curiel:2002pf,Curiel:2003uk,Arganda:2004bz}), the  $Z$-penguin contributions to the LFV observables are not suppressed by a large SUSY scale.

The paper is organised as follows: in Section~\ref{sec:mod}, we will
define the model, providing a brief overview on the implementation of
the inverse seesaw in the MSSM. In Section~\ref{sec:ZbosonLFV}, we
discuss the enhancement of the $Z$-boson mediated contributions to
low-energy cLFV observables. We also derive an analytical
approximation for the $Z-\ell_i-\ell_j$ effective vertex.  In
Section~\ref{zmlfv}, we derive analytical expressions for several
cLFV observables in the case where
$Z$-boson penguin is the dominant contribution.  In
Section~\ref{sec:res}, we detail the corresponding numerical study,
collect the relevant numerical results and discuss the results as well
as the decoupling regime. Our final remarks are given in
Section~\ref{concs}.

\section{Inverse seesaw mechanism in the MSSM}
\label{sec:mod}

The inverse seesaw model consists of a gauge singlet extension of the
MSSM. Three pairs of singlet superfields, $\widehat{\nu}^c_i$ and
$\widehat{X}_i$ ($i=1,2,3$)\footnote{We use the notation:
  $\widetilde{\nu}^c=\widetilde{\nu}_R^* $.}  with lepton numbers
assigned to be $-1$ and $+1$, respectively, are added to the
superfield content. The SUSY inverse seesaw model is defined by the
following superpotential
\begin{align}
{\mathcal W}&= \varepsilon_{ab} \left[
Y^{ij}_d \widehat{D}_i \widehat{Q}_j^b  \widehat{H}_d^a
              +Y^{ij}_{u}  \widehat{U}_i \widehat{Q}_j^a \widehat{H}_u^b 
              + Y^{ij}_e \widehat{E}_i \widehat{L}_j^b  \widehat{H}_d^a \right. \nonumber \\
              &+\left. Y^{ij}_\nu 
\widehat{\nu}^c_i \widehat{L}^a_j \widehat{H}_u^b - \mu \widehat{H}_d^a \widehat{H}_u^b \right] 
+M_{R_{ij}}\widehat{\nu}^c_i\widehat{X}_j+
\frac{1}{2}\mu_{X_{ij}}\widehat{X}_i\widehat{X}_j  ~,
\label{eq:SuperPot}
\end{align}
\noindent where $i,j = 1,2,3$ are generation indices.  In the above,
$\widehat H_d$ and $\widehat H_u$ are the down- and up-type Higgs
superfields, $\widehat L_i$ denotes the SU(2) doublet lepton
superfields. The ``Dirac''-type right-handed neutrino mass term
$M_{R_{ij}}$ conserves lepton number, while the ``Majorana'' mass term
$\mu_{X_{ij}}$ violates it by two units.  In view of this the total
lepton number $L$ is no longer conserved; notice however that in this
formulation $(-1)^L$ remains a good quantum number. Since $M_{R_{ij}}$
conserves lepton number, in the limit $\mu_{X_{ij}}\rightarrow 0$,
lepton number conservation can be restored.  In this study we consider
a general framework with three generations of $\widehat{\nu}^c$ and
$\widehat{X}$; we nevertheless recall that neutrino data can be
successfully accommodated with only one generation of
$\widehat{\nu}^c$ and $\widehat{X}$~\cite{Hirsch:2009ra}.
 
\noindent
The soft SUSY breaking Lagrangian is given by
\begin{align}
-{\mathcal L}_{\rm soft}&=-{\mathcal L}^{\rm MSSM}_{\rm soft} 
         +   \widetilde\nu^{c\dagger}_i m^2_{\widetilde \nu^c_{ij}}\widetilde\nu^c_j
         + \widetilde X^{\dagger}_i m^2_{X_{ij}} \widetilde X_j
     + (A_{\nu}^{ij} Y_\nu^{ij} \varepsilon_{ab}
                 \widetilde\nu^c_i \widetilde L^a_j H_u^b +
                B_{M_R}^{ij} M_{R_{ij}}\widetilde\nu^c_i \widetilde X_j \nonumber\\
      &+\frac{1}{2}B_{\mu_X}^{ij} \mu_{X_{ij}} \widetilde X_i \widetilde X_j
      +{\rm h.c.}),
\label{eq:softSUSY}
\end{align} where ${\mathcal L}^{\rm MSSM}_{\rm soft}$ collects the soft SUSY
breaking terms of the MSSM.   $B_{M_R}^{ij}$ and $B_{\mu_X}^{ij}$
are the new parameters involving the scalar partners of the sterile
neutrino states (notice that while the former conserves lepton number,
the latter gives rise to a lepton number violating $\Delta L=2$ term).
Assuming a flavour-blind mechanism for SUSY breaking, we consider
universal boundary conditions for the soft SUSY breaking parameters at
some very high energy scale (e.g. the gauge coupling unification scale
$\sim 10^{16}$ GeV), \beq m_\phi = m_0\,, M_\text{gaugino}= M_{1/2}\,,
A_{i}= A_0 \, \mathbb{I}\,,B_{\mu_X}=B_{M_R}= B_0 \, \mathbb{I}.  \eeq
\noindent

From Eq.~(\ref{eq:SuperPot}) one can verify that the two singlets
$\widehat{\nu}^c_i$ and $\widehat X_i$ are differently treated in the
superpotential, so that, while a $\Delta L = 2$ Majorana mass term is present
for $\widehat X_i$ ($\mu_{X_{ij}} \widehat X_i \widehat X_j$), 
no $\mu_{\nu^c_{ij}} \widehat{\nu}^c_i\widehat{\nu}^c_j$ term is included
in ${\mathcal W}$. Although the latter term can indeed be present in a
superpotential, where $(-1)^L$ is a good quantum number, we assume here for
simplicity $\mu_{\nu^c_{ij}}= 0$.  We notice that it is
the magnitude of $\mu_{X}$ (and not that of $\mu_{\nu^c}$) which
controls the size of the light neutrino
mass~\cite{Ma:2009gu,Bazzocchi:2010dt}, and that the absence of the
mass term $\mu_{\nu^c}$ does enhance the symmetry of the model (a
non-vanishing, but small value of $\mu_{\nu^c_{ij}}$ does not have
any effect on the qualitative features of the model).  Here, $\mu_{X}$
will be considered as an effective parameter, no assumption being
made on its origin. Such a parameter could be understood either
dynamically or in a SUSY Grand Unified Theory framework
\cite{Ma:2009gu,Bazzocchi:2010dt,Dev:2009aw,Dias:2012xp}. Furthermore
$\mu_{\nu^c} \ll \mu_{X}$ can also be realised in some extended
frameworks~\cite{Ma:2009gu}.

\subsection{Neutrino masses}
\label{subsec:numass}
We consider a general framework with
three generations of sterile neutrinos $\nu^c_i$ and $X_i$. 
Consequently, one has the following symmetric $(9\times9)$
mass matrix $\mathcal{M}$ in the basis
$\{\nu,{\nu^c},X\}$, 
\begin{eqnarray}
{\cal M}&=&\left(
\begin{array}{ccc}
0 & m^{T}_D & 0 \\
m_D & 0 & M_R \\
0 & M^{T}_R & \mu_X \\
\end{array}\right) \ ,
\label{nmssm-matrix}
\end{eqnarray}
Where, $m_D= \frac{1}{\sqrt 2} Y_\nu v_u$ and $M_R$, $\mu_X$ are  
$(3\times3)$ matrices in family space.
Assuming $m_D,\mu_X \ll M_R$,  the diagonalization leads to an
effective Majorana mass matrix for the light
neutrinos~\cite{GonzalezGarcia:1988rw},
\begin{equation}
\label{eqn:nu}
    m_\nu = {m_D^T M_R^{T}}^{-1} \mu_X M_R^{-1} m_D
          = \frac{v_u^2}{2} Y^T_\nu (M^T_R)^{-1} \mu_X M_R^{-1} Y_\nu.
\end{equation}
As mentioned earlier, one of the advantages of the inverse seesaw
mechanism is that the smallness of the light neutrino masses is
directly controlled by the size of $\mu_X$.  Hence the lepton number
conserving mass parameters $m_D$ and $M_R$ can easily accommodate
large (natural) Yukawa couplings ($Y_\nu\sim {\mathcal{O}}(1) $) and a
right-handed neutrino mass scale around the TeV, see
Eq.~(\ref{eqn:nu}).  In turn, this allows to have sizable
contributions to cLFV observables, contrary to what occurs in the
framework of, for example, a type-I seesaw.

In analogy to a type-I seesaw, one can define an effective right-handed
neutrino mass term $M$ such that
\begin{equation} 
\label{M-def}
M^{-1} = (M^T_R)^{-1} \ \mu_X \ M_R^{-1}.
\end{equation}
With this definition,  the light neutrino mass matrix can be cast in a way strongly resembling a standard (type-I) seesaw equation
\begin{equation} \label{M-def2}
 m_\nu = \frac{v_u^2}{2} Y^T_\nu M^{-1}  Y_\nu.
\end{equation}
This effective light neutrino mass matrix $(m_\nu)$ can be diagonalized as
\begin{equation}
    U_\text{PMNS}^T m_\nu U_\text{PMNS} = \textrm{diag}\;m_i\,.
\label{eqn:dia}
\end{equation}
Using the above equations, one can 
express the neutrino 
Yukawa couplings 
($Y_\nu = \frac{\sqrt{2}}{v_u} m_D$) as in \cite{Deppisch:2004fa} (in analogy to
to the Casas-Ibarra parameterization \cite{Casas:2001sr} in standard seesaw),
\begin{equation}
\label{c-b-iss}
Y_\nu = \frac{\sqrt{2}}{v_u} \sqrt{\hat{M}} \  R \  \sqrt{{m_i}}  \ U^\dagger_\text{PMNS},   
\end{equation}
\noindent
where $\sqrt{\hat{M}}$ is the diagonal matrix\footnote{In the subsequent sections we always consider scenarios where $M = \text{diag}(\hat{M},\hat{M},\hat{M})$. Therefore, for the sake of brevity, we use the simple notation $M$ both for the matrix and its eigenvalues.} with the square roots of the eigenvalues of $M$ and  $R$ is a $3 \times 3$ orthogonal matrix, parameterized by 3 complex angles, 
which encodes additional  mixings. 
In our study of the different cLFV observables, we assume specific regimes for $R$.

Without loss of generality, we choose a basis where $M_R$ is diagonal
at the SUSY scale, i.e.,
\begin{equation}
    M_{R_{ij}} = \textrm{diag}\;M_{R_{ii}}.
\end{equation}
In addition, in the numerical evaluation, we shall also assume
$\mu_{X_{ij}}$ to be diagonal, a simplifying assumption motivated by
the fact that cLFV observables depend only indirectly on
$\mu_{X_{ij}}$, as already discussed in the introduction. We will
further explain this issue in Section \ref{subsec:setup}. In the
subsequent analysis, we assume $M_R$, $\mu_X$ to be free parameters,
determined in order to comply with neutrino data.

Concerning neutrino oscillation data, there has been a very intense
experimental activity related to the $\theta_{13}$ mixing angle, with
very recent results from Double-Chooz~\cite{Abe:2011fz},
T2K~\cite{Hartz:2012pr}, MINOS~\cite{Adamson:2012rm},
Daya-Bay~\cite{An:2012eh} and RENO~\cite{Ahn:2012nd}. We use in our
analysis the (best-fit) results of \cite{Schwetz:2011qt}, supplemented
with the Daya-Bay measurement for $\theta_{13}$
\cite{An:2012eh}. Therefore, we consider the following set of
parameters, namely the two mass-squared differences and the three
mixing angles, given below with 3$\sigma$ range
\cite{Schwetz:2011qt,An:2012eh},
\begin{eqnarray} 
&&\sin^2\theta_{12} = 0.27-0.36 \ ,\nonumber \\ &&\sin^2\theta_{23} =0.39-0.64\ ,\nonumber \\
&&\sin^22\theta_{13} = 0.092\pm 0.06\ ,  \nonumber \\
&&\Delta m_{21}^2 = 7.09-8.19 \times 10^{-5} \; {\rm eV}^2\;,\nonumber \\ 
&&|\Delta m_{31}^2| = 2.18-2.73 \times 10^{-3}\; {\rm eV}^2\;.
\label{eq:data}
\end{eqnarray}
We also assume normal hierarchy ($\Delta m_{31}^2 > 0$). Updated
global fits to all available experimental data have also appeared
recently \cite{Tortola:2012te,Fogli:2012ua}. However, these do not
have a significant effect on our numerical results.

\section{$Z$-boson mediated cLFV}
\label{sec:ZbosonLFV}

In this section we shall discuss the enhancement of the $Z$-boson
mediated contributions,  presenting approximate results for the $Z-\ell_i-\ell_j$ effective vertex.
We then proceed to discuss how, in different extensions of the  MSSM, the $Z$-boson mediated contributions can significantly enhance several cLFV observables.
The latter effect, which is absent in the MSSM, can have a strong impact on the theoretical predictions for cLFV rates, potentially leading to a very different phenomenology since $\Br(\ell_i \to \ell_j \gamma)$
will  no longer be the most constraining observables. 

\subsection{Enhancing cLFV with the $Z$-penguin}
\label{subsec:enhancedZ}
\begin{figure}
\centering
\subfigure[0th order term: $F^{(0)}_L$]{
\includegraphics[width=0.4\linewidth]{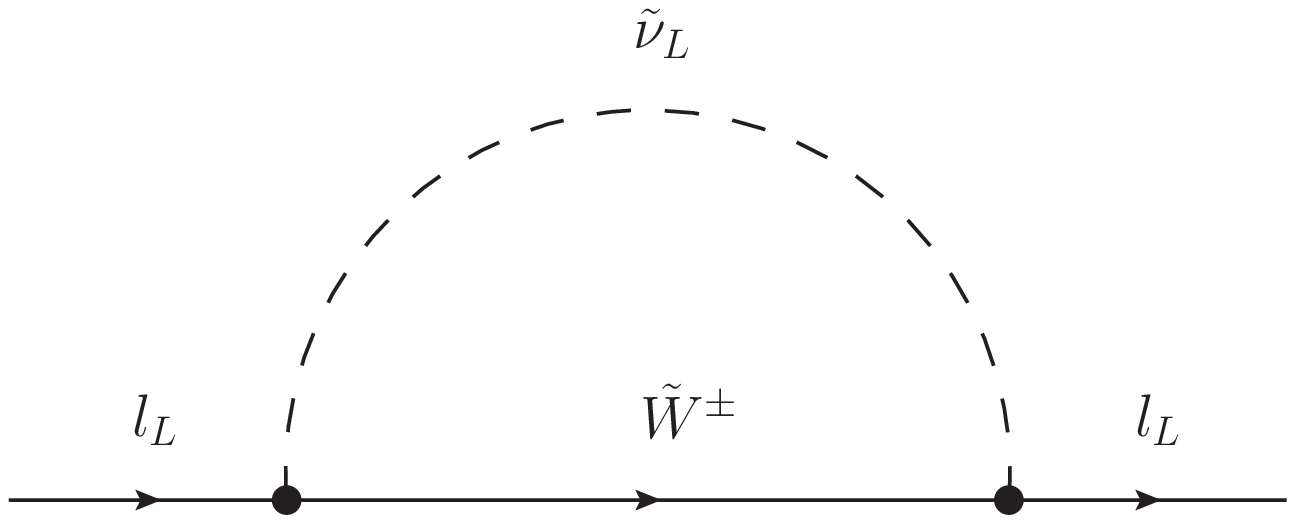}
\label{subfig:expansion0}
}
\subfigure[2nd order term: $F^{(2)}_L$]{
\includegraphics[width=0.4\linewidth]{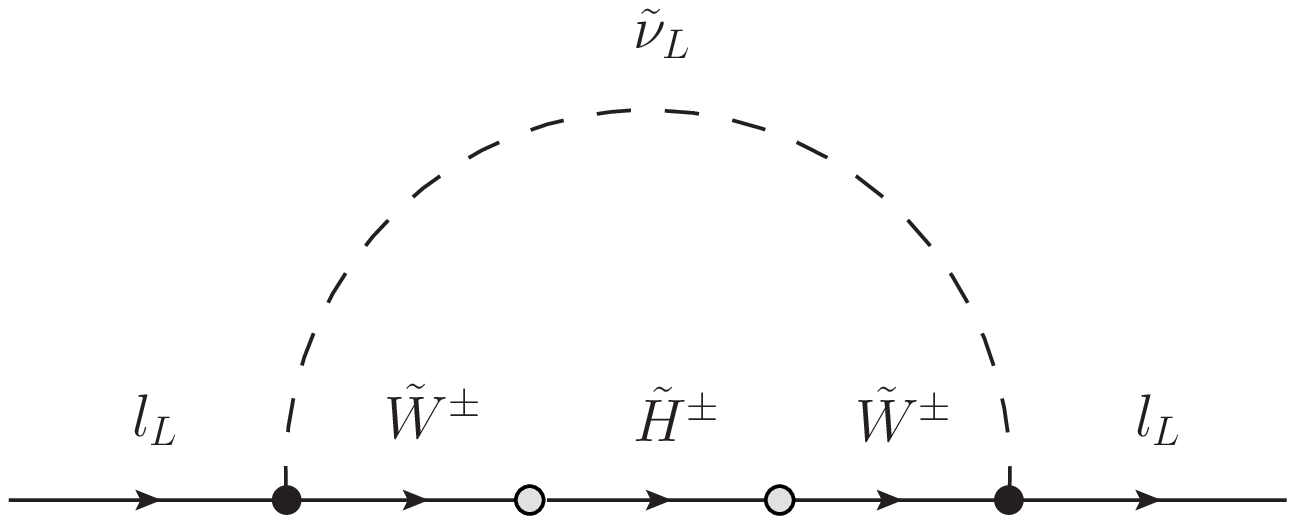}
\label{subfig:expansion2}
}
\caption{Diagrammatic representation of the $F_L$ expansion in the chargino mixing angle,  
$\theta_{\tilde{\chi}^\pm}$.}
\label{fig:expansion}
\end{figure}

In the MSSM and in many of its extensions, photon penguins provide the
dominant contributions to 3-body cLFV decays $\ell_i \to 3
\ell_j$~\cite{Hisano:1995cp,Arganda:2005ji}. The only exception arises
in the large $\tan \beta$ (and low $m_A$) regime, where large Higgs
contributions are expected \cite{Babu:2002et}\footnote{A similar
  observation in the context of the inverse seesaw was recently made
  in \cite{Abada:2011hm}.}.  However, it has recently been shown that
many simple extensions of the (s)leptonic sector can lead to large
enhancements of the $Z$-boson contributions\cite{Hirsch:2012ax},
potentially leading to scenarios where the latter provide the dominant
contribution to $\Br(\ell_i \to 3 \ell_j)$ ($> \Br(\ell_i \to \ell_j \gamma)$).
This can be understood from simple mass scaling considerations: let us
consider the chargino-sneutrino 1-loop diagrams leading to $\ell_i \to 3
\ell_j$. The photon-penguin contribution can be written as
\begin{equation} \label{Achar}
A_a^{(c)L,R} = \frac{1}{16 \pi^2 m_{\tilde{\nu}}^2} {\cal O}_{A_a}^{L,R}s(x^2)\, ,  
\end{equation}
whereas the $Z$-contributions read
\begin{equation} \label{Fchar}
F_{X} = \frac{1}{16 \pi^2 g^2 \sin^2 \theta_W m^2_{Z}}{\cal O}_{F_X}^{L,R}t(x^2)\, ,
\end{equation}
with $X=\left\{LL,LR,RL,RR\right\}$.  In the above ${\cal
  O}_{y}^{L,R}$ denote combinations of rotation matrices and coupling
constants and $s(x^2)$ and $t(x^2)$ represent the Passarino-Veltman
loop functions which depend on $x^2 =
m_{\tilde{\chi}^-}^2/m_{\tilde{\nu}}^2$ (see
\cite{Arganda:2005ji}). Notice that the only mass scale involved in
the $A$ form factors is $m_\text{SUSY}$ (the photon being massless).
On the other hand, the mass scale in the $F_X$ form factors is set, in
this case, by the $Z$-boson mass ($m_{Z}$).  Therefore, we conclude
that $A \sim m_\text{SUSY}^{-2}$ and $F \sim m_{Z}^{-2}$. Since
$m_{Z}^{2} \ll m_\text{SUSY}^2$, the $Z$-penguin can, in principle,
dominate over the photon penguin.  Assuming that all loop functions,
mixing matrices and coupling constants are of the same order, one can
estimate
\begin{equation}
\frac{F}{A} \sim \frac{m_\text{SUSY}^2}{g^2 \sin^2 \theta_W m_Z^2} \sim 500 \hspace{1cm} \text{for}  \hspace{0.5cm}\, m_\text{SUSY} = 300 \ \text{GeV}.
\end{equation}
Moreover, the $\ell_i \to 3 \ell_j$ decay width depends on $F^2$ and
$A^2$, and thus the above ratio becomes even more pronounced\footnote{
  From these considerations one can also conclude that in a very light
  SUSY scenario, with $m_\text{SUSY} \sim m_Z$, photon and
  $Z$-contributions to $\ell_i \to 3 \ell_j$ are of the same
  order. This observation has been confirmed numerically, see section
  \ref{subsec:relative}.}.  However, a subtle cancellation between the
different diagrams contributing to the leading
$Z$-contribution~\cite{Hirsch:2012ax} implies that in the MSSM, the
photon penguin is found to be (numerically)
dominant\cite{Hisano:1995cp,Arganda:2005ji}.  To understand this,
notice that the dominant contribution to $\ell_i \to 3 \ell_j$ comes
from diagrams where the leptons in the external legs are left-handed
(the other cases are suppressed by the Yukawa couplings of the charged
leptons). This is given by the form factor $F_{LL}$, usually written
as
\begin{equation}
\label{FLLfirst}
F_{LL} = \frac{F_L Z_L^{(l)}}{g^2 s_W^2 m_Z^2}\ ,
\end{equation}
where $Z_L^{(l)} = - \frac{g}{c_W}(-\frac{1}{2} + s_W^2)$ is the
$Z-\ell_i-\ell_j$ tree-level coupling ($i=j$) and $F_L$ 
is the $Z-\ell_i-\ell_j$ 1-loop effective
vertex, with $i \ne j$, 
and with $c_W = \cos
\theta_W$ and $s_W = \sin \theta_W$. $F_L$ receives  contributions from
different   1-loop diagrams, and here we   focus on the
chargino-sneutrino loop contribution. Expanding   in the chargino
mixing angle, $\theta_{\tilde{\chi}^\pm}$, one can write (see 
Fig.\ref{fig:expansion})
\begin{equation}
F_L = F^{(0)}_L + \frac{1}{2} \theta_{\tilde{\chi}^\pm}^2 F^{(2)}_L + \dots \quad .
\end{equation}
Notice that there is no term in the expansion involving $\tilde{H}^\pm$ at the
leading order, nor at the 1st order, since there is
no $\tilde{H}^\pm-\tilde{\nu}_L-\ell_L$ coupling. For this reason, only the wino contributes at the zeroth
 order in  $\theta_{\tilde{\chi}^\pm}$. 
$F^{(0)}_L$ can be written as\footnote{For the sake of clarity, we omit  flavour indices
in the expression for  $F_L$ (and in the expansion coefficients).}
\begin{equation}\label{sum-M-FL}
\left(F^{(0)}_L\right)_{ij} \equiv F^{(0)}_L = - \frac{1}{16 \pi^2} \left( M^{ij}_{\text{wave}} + M^{ij}_{\text{p1}} + M^{ij}_{\text{p2}} \right)\ , 
\end{equation}
with 
\begin{eqnarray}
M^{ij}_{\text{wave}} &=& - \frac{1}{2} g^2 (g \, c_W - g' s_W) 
                    Z_V^{xi*} Z_V^{xj} f_{\text{wave}}^x\, , \\
M^{ij}_{\text{p1}} &=& g^3 c_W Z_V^{xi*} Z_V^{xj} f_{\text{p1}}^x\, , \\
M^{ij}_{\text{p2}} &=& - \frac{1}{2} g^2 (g \, c_W + g' s_W) 
                   Z_V^{xi*} Z_V^{xj} f_{\text{p2}}^x\, ,
\end{eqnarray}
where a sum over the index $x$ is implicit. The terms in the sum come
from different types of diagrams: wave function diagrams
($M_{\text{wave}}$) and  penguins with the $Z$-boson attached to the
chargino line ($M_{\text{p1}}$) or to  the sneutrino line ($M_{\text{p2}}$). 
Moreover, $Z_V$ is a $3 \times 3$ unitary
matrix that diagonalizes the mass matrix of the sneutrinos.  Here  $f_{\text{wave}}^x = -
B_1(m_{\tilde{\chi}_1^\pm}^2,m_{\tilde{\nu}_x}^2), f_{\text{p1}}^x =
\frac{1}{2}
\tilde{C}_0(m_{\tilde{\nu}_x}^2,m_{\tilde{\chi}_1^\pm}^2,m_{\tilde{\chi}_1^\pm}^2)
- m_{\tilde{\chi}_1^\pm}^2
C_0(m_{\tilde{\nu}_x}^2,m_{\tilde{\chi}_1^\pm}^2,m_{\tilde{\chi}_1^\pm}^2),
f_{\text{p2}}^x = \frac{1}{2}
\tilde{C}_0(m_{\tilde{\chi}_1^\pm}^2,m_{\tilde{\nu}_x}^2,m_{\tilde{\nu}_x}^2)$; for the exact definitions of these loop functions,  see
\cite{Arganda:2005ji}.
The sum in Eq.~\eqref{sum-M-FL} exactly  vanishes, as can
be verified by grouping the different terms 
\begin{equation} \label{MSSM-cancellation}
F^{(0)}_L = \frac{g^2}{32 \pi^2} \left( g \, c_W Z_V^{xi*} Z_V^{xj} X_1^x 
+ g' s_W Z_V^{xi*} Z_V^{xj} X_2^x \right)
\end{equation}
with $X_1^x = f_{\text{wave}}^x - 2 f_{\text{p1}}^x + f_{\text{p2}}^x,
X_2^x = f_{\text{p2}}^x - f_{\text{wave}}^x$. Using the exact
expressions for the loop functions~\cite{Arganda:2005ji},  one finds
that the masses cancel out and that these combinations become independent of $x=m_{\tilde{\chi}^-}/m_{\tilde{\nu}}$:
 $X_1^x = X_1 = - \frac{3}{4}$ and $X_2^x = X_2 =
- \frac{1}{4}$, $\forall x$. Therefore, one is left with $F^{(0)}_L
\propto \sum_x Z_V^{xi*} Z_V^{xj} = \left( Z_V^\dagger Z_V
\right)^{ij}$, which vanishes for $i \ne j$ due to unitarity of the
$Z_V$ matrix\footnote{We notice that a similar behaviour was found in 
   \cite{Lunghi:1999uk} in association to  the $B \to X_s \ell^+
  \ell^-$ decay.}. In conclusion, the first non-vanishing term in the expansion
appears at 2nd order in the chargino mixing angle, which naturally
suppresses the $Z$-mediated contributions. This is the reason why
the photon contributions turn out to be dominant in the MSSM.

However, there are many cases where the cancellation of the zeroth
order term in the expansion no longer holds, as discussed in
\cite{Hirsch:2012ax}, where numerical examples were given, including
the inverse SUSY seesaw.  The introduction of new interactions for the
(s)leptons, in particular, the one involving the Yukawa couplings
$Y_\nu$, modifies the previous conclusion for the MSSM: owing to the
$\tilde{H}^\pm-\tilde{\nu}_R-\ell_L$ coupling, higgsino contributions
for $F^{(0)}_L$ preclude the previously discussed cancellation.  In
fact, there is a non-zero $F^{(0)}_L$ contribution that enhances $F_L$
by a huge factor leading to large $Z$-penguin mediated contributions
to cLFV processes.

It is worth mentioning that the discussed cancellation only happens for
wino diagrams and not for the higgsino ones. Therefore, $F_R$ has
already a finite zeroth order contribution in the MSSM. However, this
is very small because of the tiny charged lepton Yukawa couplings.

Finally, although the previous discussion has been focused on $\ell_i \to 3 \ell_j$,
the same enhancement in the $Z-\ell_i-\ell_j$ effective vertex
also affects other observables which are mediated by $Z$-boson exchange. 
This is the case for $\mu-e$ conversion in nuclei \cite{Arganda:2007jw} 
and $\tau \to P^0 \ell_i$, where $P^0$ is a pseudoscalar meson
\cite{Arganda:2008jj}.

\subsection{Approximate expression for the  1-loop  $Z-\ell_i-\ell_j$ effective vertex}
\label{subsec:appZLL}
This section is devoted to deriving an analytical approximation for the 1-loop  $Z-\ell_i-\ell_j$ effective vertex, $F_L \simeq F^{(0)}_L$, in the framework of the SUSY inverse seesaw. 

 Following the previous discussion, we will define this
vertex with the lepton $\ell_i$ as incoming particle. 
The
expression for $F^{(0)}_L$ obtained from the  chargino-sneutrino
loops\footnote{The analogous  neutralino-slepton loop diagrams
  provide extremely small contributions since they involve charged
  lepton Yukawa couplings and small mass insertions on the slepton
  propagators.} can be decomposed as
\begin{equation}
\label{FL0pi}
F^{(0)}_L = - \frac{1}{16 \pi^2} \left( F^{(0)}_{L,\text{wino}} + F^{(0)}_{L,\text{higgsino}} \right) \, ,
\end{equation}
where $F^{(0)}_{L,\text{wino}}$ and $F^{(0)}_{L,\text{higgsino}}$ are
the wino and higgsino contributions, respectively. The latter terms  can be rewritten as
\begin{align}
F^{(0)}_{L,\text{wino}} &= - \frac{g^2}{2} Z_V^{xi*} Z_V^{xj} \left( g c_W Y_1^x + g' s_W Y_2^x \right) \label{FL01} \\
F^{(0)}_{L,\text{higgsino}} &= \frac{g}{4 c_W} Y_\nu^{z'i *} Y_\nu^{zj} \left[ \left(c_W^2 - \frac{1}{2}\right)\frac{1}{2} \delta_{zz'} - S_{xy} Z_V^{y,z'+3 *} Z_V^{x,z+3} \tilde{C}_0 (m_{\tilde{\chi}_2^\pm}^2,m_{\tilde{\nu}_x}^2,m_{\tilde{\nu}_y}^2) \right]. \label{FL02}
\end{align}
where $Z_V$ is now a $9 \times 9$ unitary matrix and  sums over $x,y = 1,\dots,9$ and $z,z' = 1,2,3$ are implicit. We have also defined the following combinations  of loop functions:
\begin{eqnarray}
Y_1^x &=& - \tilde{C}_0(m_{\tilde{\nu}_x}^2,m_{\tilde{\chi}_1^\pm}^2,m_{\tilde{\chi}_1^\pm}^2) + 2 \ C_0(m_{\tilde{\nu}_x}^2,m_{\tilde{\chi}_1^\pm}^2,m_{\tilde{\chi}_1^\pm}^2) - B_1(m_{\tilde{\chi}_1^\pm}^2,m_{\tilde{\nu}_x}^2) \\
&& + \frac{1}{2} S_{xy} \tilde{C}_0 (m_{\tilde{\chi}_1^\pm}^2,m_{\tilde{\nu}_x}^2,m_{\tilde{\nu}_y}^2) \nonumber \\
Y_2^x &=& \frac{1}{2} S_{xy} \tilde{C}_0 (m_{\tilde{\chi}_1^\pm}^2,m_{\tilde{\nu}_x}^2,m_{\tilde{\nu}_y}^2) + B_1(m_{\tilde{\chi}_1^\pm}^2,m_{\tilde{\nu}_x}^2)
\end{eqnarray}
and
\begin{equation}
S_{xy} = \sum_{k=1}^3 Z_V^{ky *} Z_V^{kx} \, .
\end{equation}

Equations \eqref{FL01} and \eqref{FL02} are exact and do not involve any approximation nor assumption on the sneutrino mixing pattern. However, in order to render these expressions more transparent, let us now consider the following limit, 
\begin{equation}\label{Sdelta}
S_{xy} = \left\{ \begin{array}{c c}
\delta_{xy} & \text{for} \, x,y \leq 3 \\
0 & \text{for} \, x,y > 3
\end{array} \right .
\end{equation}
This assumption actually corresponds to the MSSM, since in this case $Z_V$ is a $3 \times 3$ unitary matrix. It also provides a very good approximation in the inverse seesaw when the  Yukawa couplings are relatively small. In this framework\footnote{As will be shown in our numerical analysis, regimes of small Yukawa couplings are actually  favoured by current bounds on cLFV observables. For instance, $\mu-e$ conversion in gold leads to $(Y_\nu^\dagger Y_\nu)_{12} \lesssim 3\times10^{-4}$.} the sneutrino mixing matrix $Z_V$, written in the basis $\tilde{\nu}_x = (\tilde{\nu}_{1,2,3},\tilde{\nu}^c_{1,2,3},\tilde{X}_{1,2,3})$, is approximately given by
\begin{equation} \label{ssm}
Z_V \simeq \left( \begin{array}{c | c}
\mathbb{I}_{3} & 0 \\
\hline
0 & Z_V^s
\end{array} \right)\, ,
\end{equation}
where $\mathbb{I}_{3}$ is the $3 \times 3$ identity matrix and $Z_V^s$ is a $6\times6$ unitary matrix that diagonalizes the singlet sector. The negligible left-right mixing in the sneutrino sector thus  justifies the approximation of Eq.~\eqref{Sdelta}.

In both cases, MSSM and the inverse seesaw with $\tilde{\nu}$ mixings as defined in Eq.~\eqref{ssm},  one can further simplify  Eq.~\eqref{FL01} by means of Eq.~\eqref{Sdelta}. It is straightforward to verify that $Y_1^x$ reduces to $Y_1^x = X_1 = - \frac{3}{4}$, whereas $Y_2^x$ reduces to $Y_2^x = X_2 = - \frac{1}{4}$. Therefore, using unitarity of the $Z_V$ matrix, $Z_V^{xi*} Z_V^{xj} = \delta_{ij}$, the wino contribution simplifies to
\begin{equation}
\label{FL0wino}
F^{(0)}_{L,\text{wino}} = \frac{g^2}{8} \delta_{ij} (3 g c_W + g' s_W) = \frac{g^3}{8 c_W} \delta_{ij} (1 + 2 c_W^2)\, , 
\end{equation}
which vanishes in the case of  flavour violating transitions ($i\ne j$). 
This is in perfect agreement with the $Z$-penguin cancellation in the MSSM, 
 discussed in the previous subsection.

Let us now consider the higgsino contributions $F^{(0)}_{L,\text{higgsino}}$, not present in the MSSM. In this case, the results  also  simplify under the assumption of Eq. \eqref{ssm}, which allows to cancel the last term in Eq.~\eqref{FL02}. Therefore, $F^{(0)}_{L,\text{higgsino}}$ can be written as
\begin{equation}\label{FL0final}
F^{(0)}_{L,\text{higgsino}} = \frac{g}{8 c_W} \left( Y_\nu^\dagger Y_\nu \right)_{ij} \left( c_W^2 - \frac{1}{2} \right).
\end{equation}
The above equation corresponds to the approximate LFV $Z-\ell_i-\ell_j$ 1-loop
effective vertex.

We have explicitly checked that this formula does indeed reproduce the full
numerical results to a very good approximation. Further refinements can be
obtained by including other relevant (but sub-dominant) 1-loop diagrams. In particular,
we found non-negligible sub-leading contributions arising from 1-loop diagrams
involving charged Higgs and neutrinos. We also note that $F^{(1)}_L$
and the higher order terms in the expansion receive new
contributions. However, the latter are (numerically)  negligible  when compared to
$F^{(0)}_L$.

Finally, it is worth noticing  that the
$F^{(0)}_L$ effective vertex exhibits a crucial property: at leading order there is no dependence on any supersymmetric
mass (however higher order terms in the
expansion do indeed decrease for increasing SUSY masses). 

There are well-known results in general quantum field theory
(for both SUSY and non SUSY cases) regarding the decoupling theorem
\cite{Appelquist:1974tg,Katehou:1987we}, which seems not to apply in
this case. $Z$-mediated processes exhibit a non-decoupling behaviour
and large supersymmetric masses do not suppress the charged lepton flavour
violating signatures induced by $Z$-boson exchange. This
non-decoupling behaviour has also been found in flavour changing Higgs
boson decays in the MSSM, both to hadronic
\cite{Curiel:2002pf,Curiel:2003uk} and leptonic final states
\cite{Arganda:2004bz}.

\section{Lepton flavour violating observables}
\label{zmlfv}

\subsection{Current experimental situation and future prospects}
The search for cLFV is a very active field with either dedicated experiments like MEG \cite{PhysRevLett.107.171801} or others with a broader program like $B$ factories \cite{OLeary2010af}. In this paper, we focus on leptonic observables, which  can be  classified as radiative decays, e.g. $\mu \rightarrow e \gamma$, 3-body decays, e.g. $\tau \to 3 \mu$, and neutrinoless conversion in muonic atoms, e.g. $\mu, \mathrm{Ti} \rightarrow e, \mathrm{Ti}$.

The experiments looking for radiative decays are quite different depending on the lepton in the initial state. If it is a muon, the only decay is $\mu \rightarrow e \gamma$ which is studied by dedicated experiments such as MEG \cite{PhysRevLett.107.171801}. This collaboration has plans for an upgrade that would improve the sensitivity to $\mathcal{O}(10^{-14})$. Radiative $\tau$ decays are studied at $B$ factories, which are also  $\tau$ factories, since the production cross-sections are very close at the $\Upsilon(4s)$ resonance. The current upper limits on $\mathrm{Br}(\tau \to \mu \gamma)$ and $\mathrm{Br}(\tau \to e \gamma)$ are given by the BaBar experiment, together with expected sensitivities at the future generation of $B$ factories, e.g. Belle II and SuperB \cite{OLeary2010af}.

For the same reasons 3-body decays of the $\tau$ lepton are also usually searched for at $B$ factories. The current upper limits come from the Belle experiment \cite{Hayasaka:2010np,arXiv:1011.6474} because of its larger data sample compared to BaBar. Since these observables are currently not limited by the background, significant improvements are expected at Belle II and SuperB \cite{OLeary2010af}. The decay $\mu \to 3e$  has been investigated by the SINDRUM experiment \cite{Bellgardt:1987du} and, if approved,  a future experiment named Mu3e at PSI could reach a sensitivity of $10^{-15}$ (after upgrades $10^{-16}$) \cite{LOImu3e}.

Neutrinoless conversion in muonic atoms has also been studied for different nuclei  by the SINDRUM II collaboration \cite{Dohmen:1993mp,Bertl:2006} which has set the current upper limits. In the future, the sensitivity is expected to be greatly improved by different projects\footnote{Mu2e \cite{Glenzinski:2010zz,Carey:2008zz} is a future experiment at Fermilab with expected sensitivities of respectively $10^{-17}$ (phase I) and $10^{-18}$ (phase II with Project X). On the other hand, the first experiment that could be built at J-PARC is DeeMe \cite{Aoki:2010zz} with an expected sensitivity of $2\times 10^{-14}$ in 2015. Then COMET \cite{Cui:2009zz} and 
PRISM/PRIME \cite{PRIME} would come with sensitivities of $10^{-15}$ (COMET Phase I, 2017), $10^{-17}$ (COMET phase II, 2021) and $10^{-18}$ (PRISM/PRIME) for a titanium nucleus.}. For convenience, in our numerical study we will consider a future sensitivity in the range $10^{-16}-10^{-18}$.

In addition to these low-energy observables, interesting phenomena are
expected to be observed at colliders.  One can have sizable widths for
processes like $\chi_2^0 \to \chi_1^0 \ell_i^\pm \ell_j^\mp$,
flavoured slepton mass splittings (especially between the first and
second generation of left-handed sleptons) and finally the appearance
of new edges in same-flavour dilepton mass distributions.  Assuming a
unique source of LFV (neutrino mass generation), the interplay of low-
and high-energy LFV observables can strengthen or disfavour
the underlying model of new physics.  Illustrative examples of the
potential of this interplay can be found for instance
in~\cite{Abada:2010kj,Esteves:2010si,Abada:2011mg,Abada:2012re}.

\subsection{$\mu-e$ conversion in nuclei}
This process is particularly sensitive to the enhancement of the $Z$-penguin contribution. The conversion rate, relative to the the muon capture rate, can be 
expressed as \cite{Arganda:2007jw}
\begin{align}
{\rm CR} (\mu- e, {\rm Nucleus}) &= 
\frac{p_e \, E_e \, m_\mu^3 \, G_F^2 \, \alpha^3 
\, Z_{\rm eff}^4 \, F_p^2}{8 \, \pi^2 \, Z}  \nonumber \\
&\times \left\{ \left| (Z + N) \left( g_{LV}^{(0)} + g_{LS}^{(0)} \right) + 
(Z - N) \left( g_{LV}^{(1)} + g_{LS}^{(1)} \right) \right|^2 + 
\right. \nonumber \\
& \ \ \ 
 \ \left. \,\, \left| (Z + N) \left( g_{RV}^{(0)} + g_{RS}^{(0)} \right) + 
(Z - N) \left( g_{RV}^{(1)} + g_{RS}^{(1)} \right) \right|^2 \right\} 
\frac{1}{\Gamma_{\rm capt}}\,.
\end{align}   
$Z$ and $N$ are the number of protons and neutrons in the nucleus and
$Z_{\rm eff}$ is the effective atomic charge~\cite{Chiang:1993xz}.
Similarly, $F_p$ is the nuclear matrix element and $\Gamma_{\rm capt}$
represents the total muon capture rate. $G_F$ is the Fermi constant,
$\alpha$ is the fine structure constant, $p_e$ and $E_e$ ( $\simeq
m_\mu$ in the numerical evaluation) are the momentum and energy of the
electron and $m_\mu$ is the muon mass.  In the above, $g_{XK}^{(0)}$
and $g_{XK}^{(1)}$ (with $X = L, R$ and $K = S, V$) represent the
relevant isoscalar, and isovector couplings respectively, and are
given by
\begin{align}
g_{XK}^{(0)} &= \frac{1}{2} \sum_{q = u,d,s} \left( g_{XK(q)} G_K^{(q,p)} +
g_{XK(q)} G_K^{(q,n)} \right)\,, \nonumber \\
g_{XK}^{(1)} &= \frac{1}{2} \sum_{q = u,d,s} \left( g_{XK(q)} G_K^{(q,p)} - 
g_{XK(q)} G_K^{(q,n)} \right)\,.
\end{align}
The numerical values of $G_K$ can be found in \cite{Arganda:2007jw}.

Similar to other observables involving four fermions, the $\mu-e$ conversion rate receives contributions
from $\gamma$-, $Z$- and Higgs-penguins as well as box diagrams. The corresponding couplings are,
\begin{eqnarray}
g_{LV(q)} &= &g_{LV(q)}^{\gamma} + g_{LV(q)}^{Z} + 
g_{LV(q)}^{\rm B}\,, \nonumber \\
g_{LS(q)} &=& g_{LS(q)}^{H} + g_{LV(q)}^{\rm B}\, 
\end{eqnarray}
where $g_{LX(q)}^{\gamma}$, $g_{LX(q)}^{Z}$, $g_{LS(q)}^{H}$, 
$g_{LX(q)}^{\rm B}$ (with $X=V,S$) represent the couplings 
of the photon, $Z$, $H$ and box diagrams, respectively.
Again, considering that the $Z$-boson is at the origin of the dominant 
contribution\footnote{We stress that in our numerical analysis, we will take into account all contributions to the CR. We refer the reader to \cite{Arganda:2007jw} for the corresponding formulae.}, we focus on $g_{LV(q)}^{Z}$. We thus have
\begin{eqnarray} 
g_{LV(q)} \equiv g_{LV(q)}^{Z} 
&=& -\frac{\sqrt{2}}{G_F} \, \frac{Z_L^q + Z_R^q}{2} \, \frac{F_L}{m_Z^2}\\
&\equiv&\bar{g}(q)\, \frac{F_L}{m_Z^2}\,, 
\end{eqnarray}
and
$g_{RV(q)}= g_{LV(q)}|_{L \leftrightarrow R}$. The $Z-\bar{q}-q$ couplings
($Z_L^q, Z_R^q$) can be written as 
\begin{eqnarray}
Z_L^{(q)} &=& -\frac{g}{c_W} \left[ T_3^q - Q_q s_W^2 \right], \\
Z_R^{(q)} &=& \frac{g}{c_W} Q_q s_W^2,
\end{eqnarray}
with $s_W = \sin{\theta_W}$ and $c_W = \cos{\theta_W}$.
Then, considering that  $F_R \ll F_L$, we have 
\begin{align}
{\rm CR} (\mu- e, {\rm Nucleus}) &= 
\frac{p_e \, E_e \, m_\mu^3 \, G_F^2 \, \alpha^3 
\, Z_{\rm eff}^4 \, F_p^2}{8 \, \pi^2 \, Z}  
\times \left| (Z + N)g_{LV}^{(0)}  + 
(Z - N) g_{LV}^{(1)}  \right|^2  \frac{1}{\Gamma_{\rm capt}}\, \nonumber \\
&= \frac{p_e \, E_e \, m_\mu^3 \, G_F^2 \, \alpha^3 
\, Z_{\rm eff}^4 \, F_p^2}{8 \, \pi^2 \, Z}  
\times \left| (Z + N) \bar{g}_{LV}^{(0)}  + 
(Z - N) \bar{g}_{LV}^{(1)}  \right|^2 \frac{F^2_L}{m_Z^4} 
\frac{1}{\Gamma_{\rm capt}}\,,
\end{align}   
where
\begin{align}
\bar{g}_{LV}^{(0,1)}=\frac{1}{2} \sum_{q = u,d,s} \bar{g}(q)\left(G_V^{(q,p)} \pm
 G_V^{(q,n)} \right)\,. \nonumber 
 \end{align} 
Finally, substituting $F_L$ and denoting the  hadronic coefficient 
$\left| (Z + N) \bar{g}_{LV}^{(0)}  + (Z - N) \bar{g}_{LV}^{(1)}  \right|^2 = C_\text{Had} $, we have
\begin{align}
{\rm CR} (\mu- e, {\rm Nucleus}) 
= \frac{p_e \, E_e \, m_\mu^3 \, G_F^2 \, \alpha^3 
\, Z_{\rm eff}^4 \, F_p^2 g^2}{2^{17} \, \pi^4 \, Z \,c_W^2}  
\times C_\text{Had} \frac{(c^2_W-\frac{1}{2})^2}{m_Z^4}\,
\left( Y_\nu^\dagger Y_\nu \right)^2_{12} 
\frac{1}{\Gamma_{\rm capt}}\,.
\end{align}   

\noindent This approximate formula is valid for   scenarios where
$Z$-boson provides the dominant contribution to the conversion rate. 

\subsection{$\Br(\ell_i \to 3 \ell_j)$}
As we discussed in Section~\ref{sec:ZbosonLFV},  the $Z$-penguin  can provide the dominant contribution to the decay width of $\ell_i \to 3 \ell_j$ in the inverse seesaw extension of 
MSSM. In the limit of $Z$-penguin dominance, the decay width 
can be cast as \cite{Hisano:1995cp,Arganda:2005ji}
\begin{align}
\Gamma(\ell_i \to 3 \ell_j) & \simeq \frac{e^4}{512\pi^3} m^5_{\ell_i} \frac{2}{3}
(F^2_{LL}+F^2_{RR}+F^2_{LR}+F^2_{RL})\\ \nonumber
&\simeq \frac{e^4}{512\pi^3} m^5_{\ell_i} \frac{2}{3} F^2_{LL}.
\end{align}
The complete expression for $F_{LL}$ can be derived using 
Eqs. \eqref{FLLfirst},\eqref{FL0pi},\eqref{FL0wino} and \eqref{FL0final},
\begin{align}
F_{LL} & = - \frac{(\frac{1}{2}-s^2_W)}{128 \pi^2 s^2_W c^2_W m_Z^2}\left[
g^2 \delta_{ij} (1 + 2 c_W^2)+\left( Y_\nu^\dagger Y_\nu \right)_{ij} \left( c_W^2 - \frac{1}{2} \right)\right]\\
& = - \frac{(\frac{1}{2}-s^2_W)^2}{128 \pi^2 s^2_W c^2_W m_Z^2}\left( Y_\nu^\dagger Y_\nu \right)_{ij} \ \quad  (i\ne j), 
\end{align}
so that the
decay width for $\Gamma(\ell_i \to 3 \ell_j)$ is given by 
\begin{align}
\Gamma(\ell_i \to 3 \ell_j) 
&= \frac{e^4(\frac{1}{2}-s^2_W)^4}{3\cdot 2^{22} \pi^7 s^4_W c^4_W m_Z^4} m^5_{\ell_i} 
\left( Y_\nu^\dagger Y_\nu \right)^2_{ij}.
\end{align}
In this limit one recovers a correlation between 
CR$(\mu-e , {\rm Nucleus})$ and $\Br(\mu \to 3e)$,  which we shall briefly mention in our
numerical results.

\subsection{$\Br(\tau \to P^0 \ell_i)$}
$\tau$ decays into a light charged lepton and a pseudoscalar
meson\footnote{We do not consider final states with a $K^0$ as this
  would involve additional suppression due to the double penguin.},
$P^0 = \pi^0,\eta,\eta'$, can also be significantly enhanced in this
framework, owing to $Z$-boson mediation.  The corresponding branching
ratio can be written as \cite{Arganda:2008jj}
\begin{equation}
  \Br(\tau \to P^0 \ell_i) \, = \, \frac{1}{4 \pi} \,
  \frac{\lambda^{1/2}(m_\tau^2,m_{\ell_i}^2,m_P^2)}{m_\tau^2 \,\,\Gamma_\tau} \,
\frac{1}{2} \,  \sum |T|^2 \, ,
\end{equation}
where  $\lambda(x,y,z)= (x+y-z)^2-4xy$ and $\Gamma_\tau$
is the total decay width of the $\tau$ lepton. 
The averaged squared amplitude summed over initial and final states 
is given by   
\begin{equation}
 \frac{1}{2} \sum |T|^2  =  \frac{1}{4 \, m_\tau} 
\sum_{k,m} \left[ 2 m_{\ell_i} m_\tau \left( a^k_P a^{m\,*}_P - b^k_P b^{m\,*}_P \right) 
+ (m_\tau^2 + m_{\ell_i}^2 - m_P^2 ) \left( a^k_P a^{m\,*}_P  + b^k_P b^{m\,*}_P \right)
\right],
\end{equation}
with $k,m = Z,A^0$. Focusing on the $Z$-boson contributions,   we have
\begin{eqnarray}
 a^Z_P & = & - \frac{g}{2 \cos\theta_W} \frac{F_\pi}{2} \frac{C(P)}{m_Z^2} 
 \left( m_\tau-m_{\ell_i} \right) 
\left( F_L + F_R \right) \, ,\nonumber \\
b^Z_P & = &  \frac{g}{2 \cos\theta_W} \frac{F_\pi}{2} \frac{C(P)}{m_Z^2} 
\left( m_\tau+ m_{\ell_i} \right) 
\left( F_R - F_L \right) \, , 
\end{eqnarray}
where, $F_R = F_L |_{L \leftrightarrow R}$, which is in general small. In the
above, $F_\pi~\simeq~92.4$ MeV is the pion decay constant while the
functions $C(P)$ have been defined in \cite{Arganda:2008jj}.
In the limit $F_R \ll F_L$, we have
\begin{equation}
 \frac{1}{2} \sum |T|^2  =  
\frac{1}{256 \, m_\tau} \frac{g^2F_\pi^2C(P)^2}
{c^2_W m_Z^4}\left[ -8m^2_\tau m^2_{\ell_i}+(m_\tau^2 + m_{\ell_i}^2 - m_P^2 )(m_\tau^2 + m_{\ell_i}^2)\right]F^2_L\ .
\end{equation}
Finally, using the expression for $F_L$, one finds
\begin{align}
 \frac{1}{2} \sum |T|^2  & =  
\frac{1}{2^{19}\, m_\tau} \frac{g^4F_\pi^2C(P)^2}
{\pi^4 c^4_W m_Z^4}\left[ -8m^2_\tau m^2_{\ell_i}+(m_\tau^2 + m_{\ell_i}^2 - m_P^2 )(m_\tau^2 + m_{\ell_i}^2)\right]\nonumber \\ 
& \times (c^2_W-\frac{1}{2})^2\left( Y_\nu^\dagger Y_\nu \right)^2_{i3}.
\end{align}
Although this approximate formula corresponds to leading order estimates, 
in the numerical evaluation, we take into account all contributions 
from the $Z$- and Higgs bosons.

\subsection {$\Br(\ell_i \to \ell_j \gamma)$}
\label{obs:meg}
For completeness, we discuss the radiative decays $\ell_i \to \ell_j \gamma$. These observables, which have been studied in great detail in~\cite{Hisano:1995cp,Hisano:1995nq,Deppisch:2004fa},  are not sensitive to the $Z$-mediation. 
It is however known 
that, for small values of $M_R$, the branching fraction can easily reach 
the experimental sensitivity, even in the absence of supersymmetric contributions \cite{Deppisch:2004fa}. 
In the following section, considering both heavy
singlet neutrino as well as  supersymmetric contributions, we will explore 
the part of the parameter space where $\Br(\ell_i \to \ell_j \gamma)$ is within experimental sensitivity reach.

\subsection{Collider observables: $\Br(\tilde{\chi}_2^0 \to \tilde{\chi}_1^0 \ell_i \ell_j)$ and $\Delta m_{\tilde{\ell}}$}\label{coll}

\begin{figure}
\centering
\includegraphics[width=0.4\linewidth]{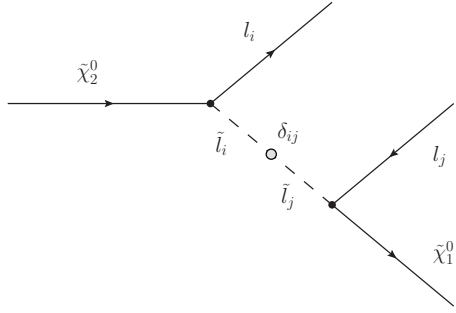}
\caption{Feynman diagram contributing to 
$\Br(\tilde{\chi}_2^0 \to \tilde{\chi}_1^0 \ell_i \ell_j)$. The white white circle represents a LFV mass insertion in the charged slepton propagator.}
\label{Xi02toXi01}
\end{figure}

So far, we have only addressed low-energy observables.  However, cLFV
collider observables such as $\Br(\tilde{\chi}_2^0 \to
\tilde{\chi}_1^0 \ell_i \ell_j)$ and the slepton mass splittings
$\Delta m_{\tilde{\ell}}$ ($m_{\tilde{\ell_i}}-m_{\tilde{\ell_j}}$) are also
relevant in scenarios that provide a strong connection between low-
and high-energy lepton flavour violation
\cite{Abada:2010kj,Esteves:2010si,Abada:2011mg,Abada:2012re}.

However, in the model under investigation, one expects that these
high-energy observables will be of little relevance when compared to
the low-energy cLFV observables. For example, the decay
$\tilde{\chi}_2^0 \to \tilde{\chi}_1^0 \ell_i \ell_j$ is induced by
diagrams like the one shown in Figure \ref{Xi02toXi01}. As we will
verify in the following section, the experimental limits on the
low-energy observables require small neutrino Yukawa couplings
(typically of order $\sim 10^{-3}-10^{-2}$), which in turn leads to low $\tilde{\chi}_2^0 \to \tilde{\chi}_1^0 \ell_i \ell_j$ rates and 
small mass splittings.

For completeness, we have also studied LFV decays such,  as $Z \to \ell_i\ell_j$ and $h^0 \to \ell_i \ell_j$. However, as explained below, no enhancement was found.

\section{Numerical results}
\label{sec:res}

In this section we present our numerical results. We begin by introducing
the basic setup for our computation and show some results concerning
the relative importance of the different contributions to the cLFV
observables. We then discuss our results for the rates of cLFV 
observables, emphasising the most relevant features of this model. 
Finally, we provide a brief summary of the main results, which we illustrate via  some representative benchmark points.

\subsection{Setup}
\label{subsec:setup}
Our numerical results have been obtained with a {\tt SPheno} \cite{Porod:2003um,Porod:2011nf} code generated with the Mathematica package {\tt SARAH} \cite{Staub:2011dp,Staub:2010jh,Staub:2009bi}. The computation of the LFV observables is based on the results presented in~\cite{Arganda:2005ji,Arganda:2007jw,Arganda:2008jj}, which we extended to the inverse seesaw case. 

In what concerns the supersymmetric parameters, the model is characterised by
\beq
 m_0, M_{1/2}, A_0, \tan\beta, \text{sign}(\mu), B_0.
 \eeq
Due to their little influence on the relevant observables, 
we fix $\text{sign}(\mu)$ ($+$) and $B_0(=0)$ in our scans\footnote{It is worth recalling that $B_0$ only affects the sneutrino sector. $\mu$ and $B_\mu$ are computed using the tadpole equations at the SUSY scale.}. We have explicitly checked that only $\mu \to e \gamma$ depends on the $B_0$ parameter. This can be  understood from its effect on the sneutrino spectrum, which in turn strongly affects $\Br(\mu \to e \gamma)$. We find that in some cases one can enhance $\Br(\mu \to e \gamma)$ by as much as one order of magnitude if one uses $B_0$ to fine-tune the superparticle masses. The other  observables are nearly independent of $B_0$.  

Concerning the remaining cMSSM-like parameters, $m_0$, $M_{1/2}$,
$A_0$ and $\tan \beta$, we will consider some specific scenarios. We
nevertheless recall that the $Z$-penguin, dominant in most of the
parameter space, has a very mild dependence on the cMSSM parameters,
so that our conclusions for the $Z$-boson mediated processes are quite
general.  On the other hand, for the radiative $\ell_i \to \ell_j
\gamma$, the dependence on the SUSY parameters cannot be
neglected. These observables have been studied in SUSY and non-SUSY
scenarios
\cite{Deppisch:2004fa,Hirsch:2009ra,Abdallah:2011ew,LalAwasthi:2011aa}
and we include them in our analysis for completeness. As we will see,
only for very low values of $M_R$ or when the superparticles in the
loop are light, they become the most constraining observables.

As explained in Section \ref{subsec:numass}, we will work in the basis where $M_R$ is a diagonal matrix and compute the resulting  Yukawa couplings  by means of the Casas-Ibarra parameterisation \cite{Casas:2001sr}. In this basis, both $Y_\nu$ and $\mu_X$ can be in general non-diagonal. In fact, neutrino mixing requires off-diagonal entries in at least one of these two matrices. In principle, one can consider two limits: (1) \emph{$Y_\nu$ limit}: $Y_\nu$ contains off-diagonal entries and $\mu_X$ is diagonal, and (2) \emph{$\mu_X$ limit}: $Y_\nu$ is diagonal and $\mu_X$ contains off-diagonal entries (any phenomenologically allowed scenario will be either $Y_\nu$ limit, $\mu_X$ limit or an intermediate case). However, the off-diagonal elements in $\mu_X$ have no impact on the phenomenology in the charged lepton sector, since this parameter does not appear in the coefficients of the operators relevant for the cLFV transitions. Therefore, in the $\mu_X$ limit (where $Y_\nu$
  is diagonal),  all cLFV observables
vanish. Thus, we choose to work in the $Y_\nu$ limit, assuming $\mu_X = \hat{\mu}_X \, \mathbb{I}$, where $\mathbb{I}$ is the $3 \times 3$ identity matrix, which 
in turn maximises the lepton flavour violating effects allowed by this model.
However, we stress that the model cannot be ruled out by the non-observation of cLFV processes, since one can always approach the $\mu_X$ limit to suppress the corresponding cLFV observables. 

Concerning the $M_R$ matrix, we will consider two scenarios: degenerate and non-degenerate spectrum, for simplicity presenting our results for the first one.
In any case, and as shown below, the $Z$-boson mediated processes have very little dependence on the right-handed (s)neutrino spectrum. We further choose $R = 1$ (see Eq.~(\ref{c-b-iss})) in order to keep the discussion as simple as possible\footnote{In general, the limit $R=1$
translates into a ``conservative'' limit for flavour violation:
apart from possible cancellations, and for a fixed
seesaw scale, this limit typically provides
a lower bound on the generated cLFV rates.}. 

Let us now discuss the relevant parameters for the cLFV observables. As explained above, $\mu_X$ has very little impact on the phenomenology, since it only affects LFV due to its relation to the size of  $Y_\nu$. On the other hand, $M_R$ has a stronger impact, since it affects the masses of the singlet states, but,  as already explained, the $Z$-penguins are only slightly sensitive to the spectrum. Therefore, the impact of $M_R$ on cLFV is rather small, only via the size of the  Yukawa couplings. The relevant quantities for the analysis are $M \sim \frac{M_R^2}{\mu_X}$ (as defined in Eq. \eqref{M-def}), controlling the size of the Yukawa couplings $Y_\nu$, and $M_R$, which plays a sub-dominant r\^ole
\footnote{The exception to this rule is $\mu \to e \gamma$: 
 as one lowers $M_R$, one enhances $\text{Br}(\mu \to e \gamma)$; 
this is due to non-SUSY contributions, see Section \ref{obs:meg}. Such  an effect  is less pronounced in other observables.}. Therefore, we will study the variation of the cLFV rates with respect to $M$, showing how different possibilities for $M_R$ affect the numerical results.

In this work, we do not take into account the LHC Higgs mass
 constraint on
the allowed parameter space. 
Our conclusions for lepton flavour violating observables are expected to hold 
when the Higgs mass constraint is forced upon the cMSSM parameter space.  Indeed, one can always find regions in the cMSSM
parameter space that fulfill the recent results from the LHC
\cite{ATLAS:2012ae,Chatrchyan:2012tx} and Tevatron
\cite{TEVNPH:2012ab}, pointing towards a Higgs mass in the 125 GeV range:  this can be achieved with large $A_0$ (to increase the mixing in
the stop sector) and a moderately large $\tan \beta$, see for example
\cite{Arbey:2011ab,Baer:2011ab,Ellis:2012aa}. This conclusion can easily be generalized to the
supersymmetric inverse seesaw. 
 
 In the limit of small Yukawa couplings, the singlet sector contributions to the Higgs mass are negligible~\cite{Elsayed:2011de}.
 In fact, and as we will see below, limits on cLFV processes set important constraints on the Yukawa couplings, forcing them to be below $Y_\nu \sim 10^{-3}-10^{-2}$ (unless one is working in the $\mu_X$ limit defined
above). 
 Thus one simply recovers the cMSSM
result. We will discuss a particular benchmark point
where this is explicitly shown.

\subsection{Relative importance of the different contributions}
\label{subsec:relative}

We start this section by considering the different contributions of the SM/SUSY particles
to the cLFV observables. We  display our results for $\ell_i \to 3 \ell_j$ and $\mu-e$ conversion in gold, although the conclusions drawn here can be easily extended to other cLFV observables.

\begin{figure}
\centering
\subfigure[$\, \mu \to 3 e$]{
\includegraphics[width=0.45\linewidth]{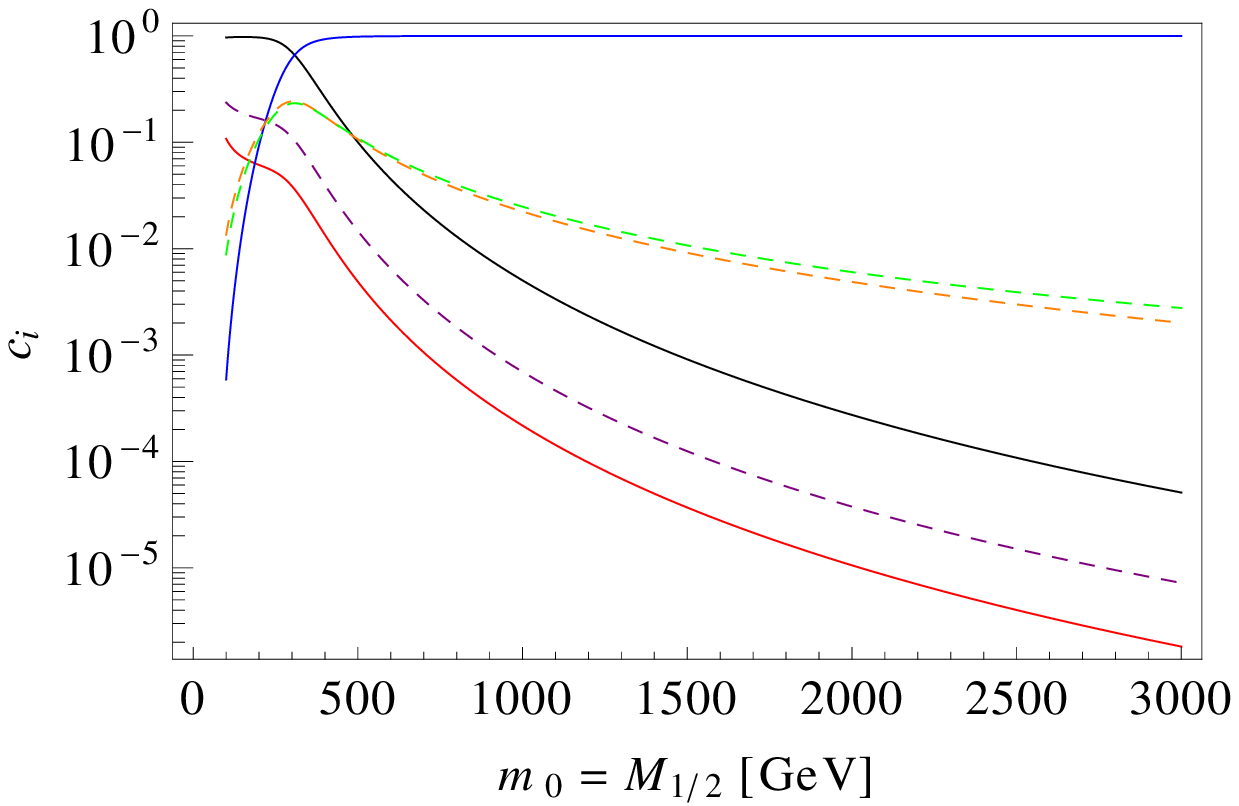}
\label{contMEEE}
}
\subfigure[$\, \mu-e$ conversion in $^{197}_{79}\text{Au}$]{
\includegraphics[width=0.45\linewidth]{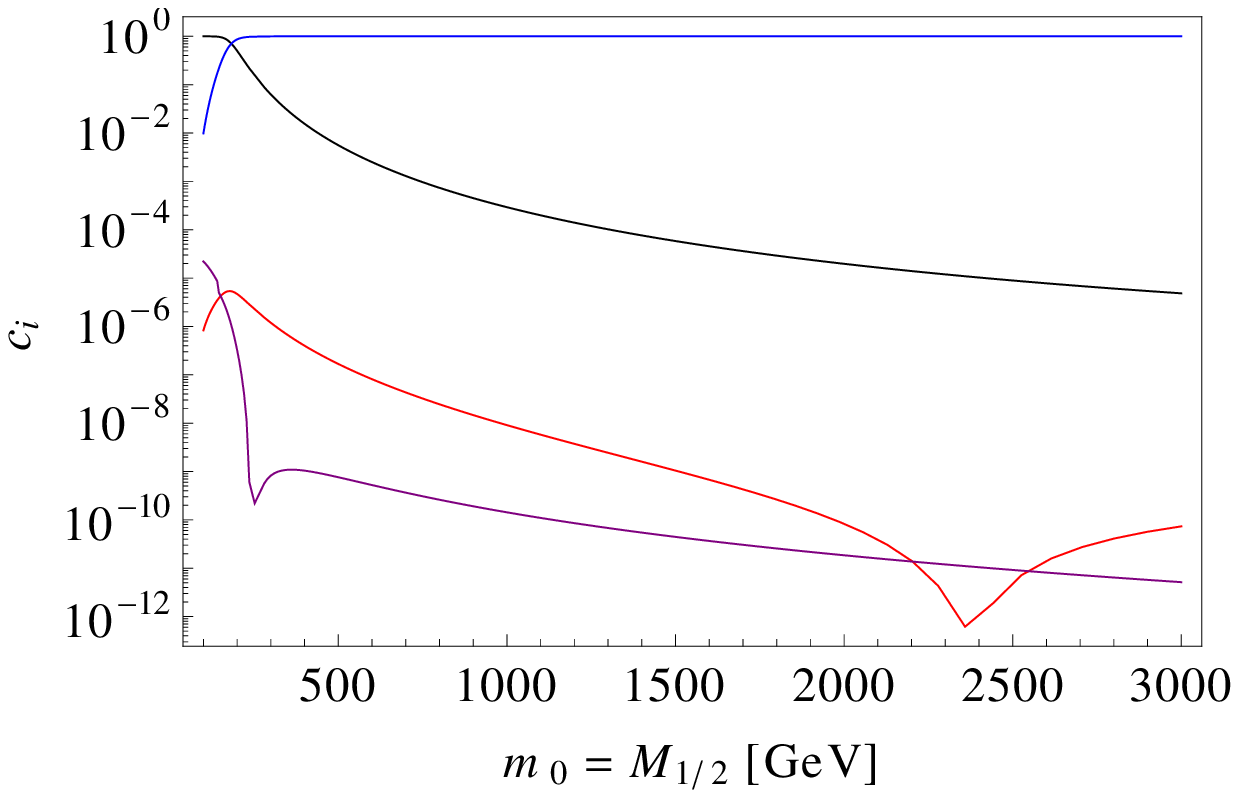}
\label{contMUE}
}
\caption{Relative contributions to $\Br(\mu \to 3 e)$ (left-hand side) and $\mu-e$ conversion in $^{197}_{79}\text{Au}$ (right-hand side) as a function of $m_0 = M_{1/2}$ for a degenerate singlet spectrum with $\hat{M}_R = 10$ TeV and $M = 10^{11}$ GeV. The rest of the cMSSM parameters are set to  $A_0 = -300$ GeV, $B_0 = 0$, $\tan \beta = 10$ and $\text{sign}(\mu)=+$. On the left-hand side, solid lines represent individual contributions, $\gamma$ (black), $Z$ (blue) and $h$ (red) whereas the dashed lines represent interference terms, $\gamma-Z$ (green), $\gamma-h$ (purple) and $Z-h$ (orange). Note that in this case $h$ includes both Higgs and box contributions. On the right-hand side interference terms are not shown to make the results clearer. The individual contributions are $\gamma$ (black), $Z$ (blue) and $h$ (red) and boxes (purple).}
\label{fig:contributions}
\end{figure}

Figure \ref{fig:contributions} shows how the different relative contributions to $\Br(\mu \to 3 e)$ and $\mu-e$ conversion in $^{197}_{79}\text{Au}$ depend on the SUSY scale, setting for simplicity  $m_0 = M_{1/2}$. The relative contributions, denoted as $c_i$, are defined as
\begin{equation} \label{def-ci}
c_i = \frac{|\text{R}_i|}{\sqrt{\sum_k \text{R}_k^2}},
\end{equation}
where $\text{R}_i$ is the rate (branching ratio in case of $\mu \to 3 e$ and conversion rate in case of $\mu-e$ conversion in Au) that would be obtained should the $i$-contribution to the process be the only one. These numerical results have been obtained with fixed values for $A_0 = -300$ GeV, $B_0 = 0$, $\tan \beta = 10$ and $\text{sign}(\mu)=+$. Nevertheless, we have verified that a different set of parameters would lead to similar results.

For $\Br(\mu \to 3 e)$ the different contributions include $(\gamma, Z, h, \text{interference} \, \gamma-Z, \text{interference} \, \gamma-h, \text{interference} \, Z-h)$. Note that $h$ includes both Higgs and box contributions, grouped together in  \cite{Arganda:2005ji}. Figure \ref{contMEEE} clearly shows that the $Z$-boson contribution to the process is the dominant one. Only for very low $m_0 = M_{1/2}$ can one find competitive (even dominant) 
photon contributions, as expected from the theoretical arguments, see Section \ref{sec:ZbosonLFV}. However, the low $m_0$ region has  already been excluded by direct collider searches \cite{Aad:2011ib,Chatrchyan:2011zy}.

A very similar behaviour is found for $\mu-e$ conversion in Au, where
the $Z$-boson contribution turns out to be the dominant one as
well. Figure \ref{contMUE} depicts the different individual
contributions to the conversion rate, with $i = (\gamma, Z, h,
\text{boxes})$. In this case, interference terms have not been shown
to simplify the plot and to render the main results more visible.
Again, as occurred for $\Br(\mu \to 3 e)$, the photon contribution
becomes important for light SUSY scenarios (lower values of $m_0 =
M_{1/2}$).  The little dip in the boxes curve is due to small
numerical instabilities in the loop function computation.  The
reduction of the photon contribution in $\mu-e$ conversion, when
compared to the $\Br(\mu \to 3 e)$, is due to the smaller electric
charges of the quarks compared to those of the leptons.

\begin{figure}
\centering
\includegraphics[width=0.5\linewidth]{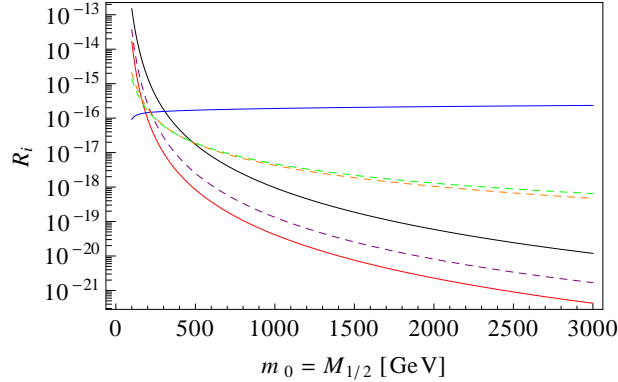}
\caption{Absolute contributions to $\Br(\mu \to 3 e)$ as a function of $m_0 = M_{1/2}$. The parameters and the color code are the same as in figure \ref{fig:contributions}.}
\label{fig:contributions2}
\end{figure}

Finally, we  show in Figure \ref{fig:contributions2} the absolute individual contributions to $\Br(\mu \to 3 e)$ as a function of $m_0 = M_{1/2}$. This plot, complementary to the one on the left-hand side of Figure \ref{fig:contributions}, serves to illustrate the non-decoupling behaviour of the $Z$-boson contributions. As one increases $m_0 = M_{1/2}$, the resulting SUSY spectrum becomes heavier and the photon, Higgs and box contributions to $\Br(\mu \to 3 e)$  clearly decrease. In contrast, the $Z$-penguin contribution remains constant in the heavy SUSY limit, in which case the 1-loop effective $Z-\mu-e$ coupling can be well described by the leading order term $F_L^{(0)}$, see Eq. \eqref{FL0final}.

In conclusion, in the framework of the inverse seesaw, should the
$Z$-penguin be present, it will provide the dominant contribution to
cLFV observables (except in the case of a very light SUSY spectrum).
This result clearly justifies the approximations of Section
\ref{zmlfv}.

\subsection{$\mu-e$ conversion in nuclei}
\label{subsec:mueresults}

After the discussion on the relative  size of the different individual contributions, we proceed to study the different cLFV observables and  begin with $\mu-e$ conversion rates  in nuclei\footnote{This has been studied in \cite{Deppisch:2005zm} for the non-supersymmetric inverse seesaw case.}.

We first address the degenerate singlet scenario, $M_R = \text{diag}(\hat{M}_R,\hat{M}_R,\hat{M}_R)$.  The more general case of a hierarchical spectrum will be briefly discussed in Section \ref{res:summ}, although we should mention that the difference for this observable is small and does not affect our conclusions.

\begin{figure}
\centering
\subfigure[$^{197}_{79}\text{Au}$]{
\includegraphics[width=0.45\linewidth]{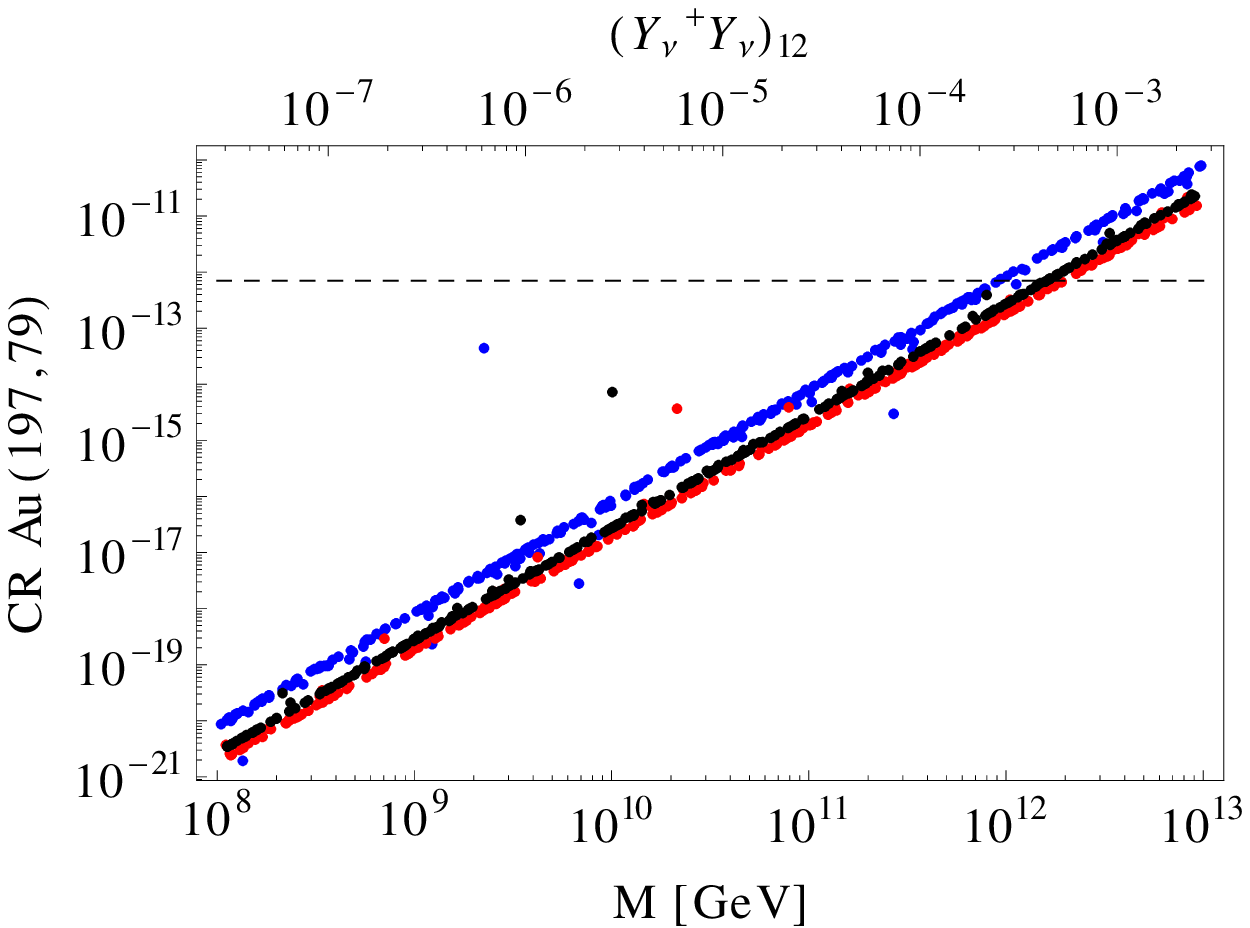}
\label{CRAu-scatter}
}
\subfigure[$^{48}_{22}\text{Ti}$]{
\includegraphics[width=0.45\linewidth]{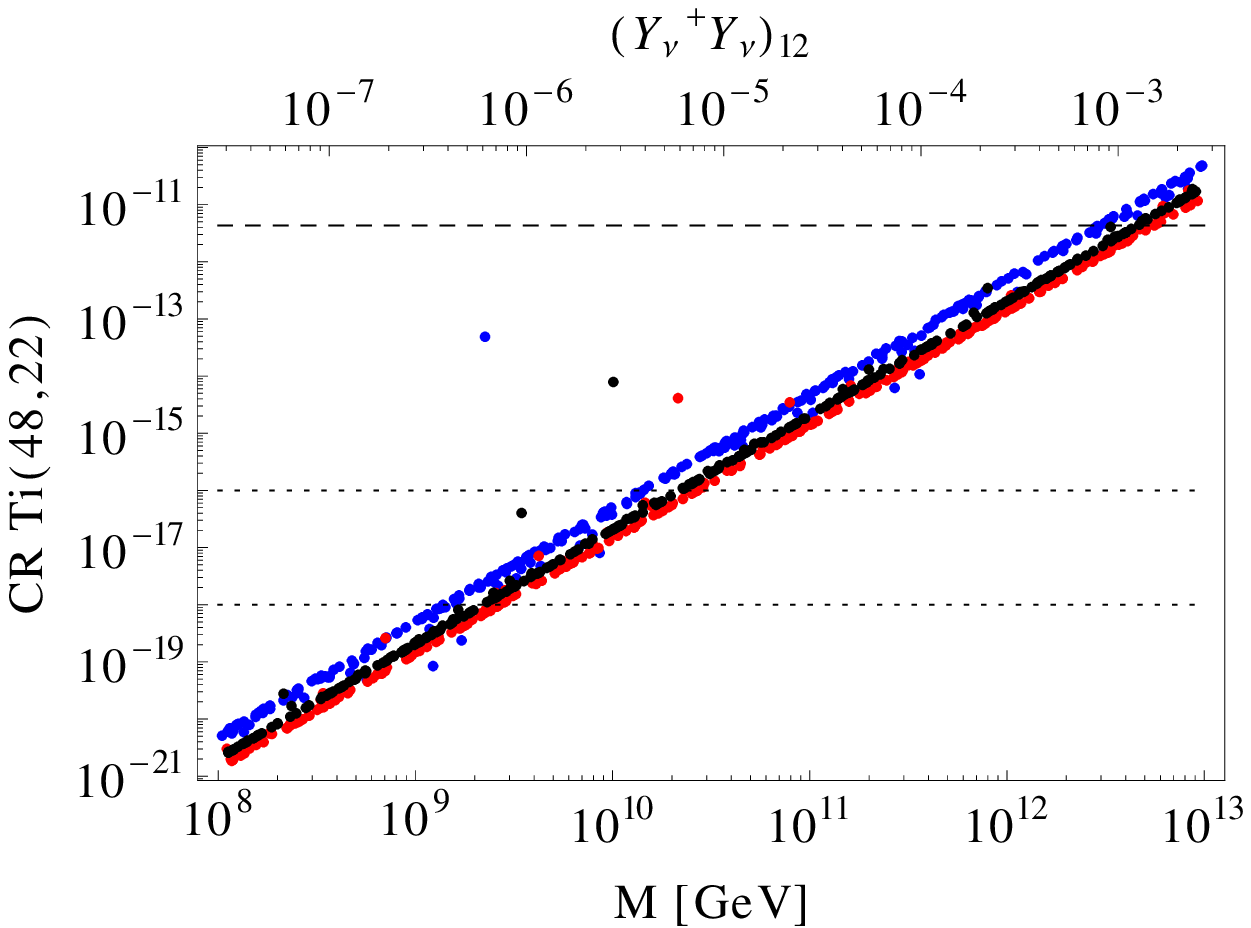}
\label{CRTi-scatter}
}
\caption{$\mu-e$ conversion rates in $^{197}_{79}\text{Au}$  (left) and $^{48}_{22}\text{Ti}$ (right), as a function of $M$ and $(Y_\nu^\dagger Y_\nu)_{12}$ for different $\hat{M}_R$ values: $\hat{M}_R = 100$ GeV (blue), $\hat{M}_R = 1$ TeV (red) and $\hat{M}_R = 10$ TeV (black). We set $A_0 = -300$ GeV, $\tan \beta = 10$, $\text{sign}(\mu) = +$ and $B_0 = 0$,  and we randomly  vary $m_0$ and $M_{1/2}$  in the range [$0,3$ TeV]. The horizontal dashed lines represent the current experimental bounds and the dotted ones represent the expected future sensitivities.}
\label{CRfig}
\end{figure}

Figure \ref{CRfig} shows the $\mu-e$ conversion rates for different nuclei $^{197}_{79}\text{Au}$ (left) and $^{48}_{22}\text{Ti}$ (right), as a function of $M$ for three different $\hat{M}_R$ values ($100$ GeV, $1$ TeV, $10$ TeV). Again, we take $A_0 = -300$ GeV, $\tan \beta = 10$, $\text{sign}(\mu) = +$ and $B_0 = 0$, but $m_0$ and $M_{1/2}$ are randomly taken in the range [$0,3$ TeV]. Note that there is a sharp correlation with $M$, hardly distorted 
   by the changes in $m_0$ and $M_{1/2}$. For the red and black dots, 
this is a consequence of $Z$-boson 
dominance, as discussed in Section \ref{sec:ZbosonLFV}. 
 The blue dots correspond to the limiting case of non-SUSY photon-penguin dominance (associated to $\hat{M}_R = 100$ GeV), which leads to larger values for $\mu-e$ conversion rates.

\begin{table}[hbt]
\centering
\begin{tabular}{c c c}
\hline
 & $M$ [GeV] & $(Y_\nu^\dagger Y_\nu)_{12}$ \\
\hline
$\text{CR}_{\text{Au}}$(current) & $10^{12}$ & $2.7  \times 10^{-4}$ \\
$\text{CR}_{\text{Ti}}$(current) & $4  \times 10^{12}$ & $10^{-3}$ \\
$\text{CR}_{\text{Ti}}$(future: $10^{-16}$) & $2  \times10^{10}$ & $5.5 \times10^{-6}$ \\
$\text{CR}_{\text{Ti}}$(future: $10^{-18}$) & $2  \times 10^{9}$ & $5.5 \times10^{-7}$ \\
\hline
\end{tabular} 
\caption{Approximate limits on $M$ and $(Y_\nu^\dagger Y_\nu)_{12}$ from the non-observation of $\mu-e$ conversion in nuclei. This table includes current experimental bounds and future expected sensitivities~\cite{Glenzinski:2010zz,Cui:2009zz,Carey:2008zz}.}
\label{tab:muebounds}
\end{table}

The non-observation of $\mu-e$ conversion in gold,  $^{197}_{79}\text{Au}$, sets an upper bound on the size of $M$,  $M \sim 10^{12}$ GeV (the exact value slightly dependent on $\hat{M}_R$).

Conservative limits will be derived from the results obtained for $\hat{M}_R \gtrsim 1$ TeV. We stress that  lower values for $\hat{M}_R$ would lead to more stringent limits. This approximate limit on $M$ can be translated  into constraints for  $(Y_\nu^\dagger Y_\nu)_{12}$, the flavour violating combination of Yukawa couplings controlling  $\mu-e$ transitions. Using 
Eqs. (\ref{M-def}) and (\ref{c-b-iss}), 
we find that this limit is $2.7 \times 10^{-4}$. On the other hand, $\mu-e$ conversion in $^{48}_{22}\text{Ti}$ has a slightly more relaxed experimental bound, which in turn implies less stringent limits on both $M$ and $(Y_\nu^\dagger Y_\nu)_{12}$, $ 4 \times 10^{12}$ GeV and $10^{-3}$,  respectively. However, there are 
plans to improve the experimental sensitivities  for $\mu-e$ conversion in titanium  with expected sensitivities in the $10^{-18}-10^{-16}$ range~\cite{Glenzinski:2010zz,Cui:2009zz,Carey:2008zz}. Should this be the case, and no $\mu-e$ conversion events be observed, the limits on the parameters of the inverse seesaw model would be greatly improved. 

In Table \ref{tab:muebounds} we compute approximate limits for $M$ and $(Y_\nu^\dagger Y_\nu)_{12}$, as obtained from current experimental bounds and future expected sensitivities.
Such  a (rough) estimate is possible due to the little dependence of this result on  $\hat{M}_R$ and on the cMSSM parameters, as shown in Figure \ref{CRfig}. 
However, from these estimates, we can safely conclude that $\mathcal{O}(1)$ Yukawa couplings, as typically chosen in the literature, are clearly ruled out since they would induce excessively  large cLFV effects. This would imply that $(Y_\nu^\dagger Y_\nu)_{12} \sim 1$, corresponding  to $M \sim 10^{14}$ GeV, 
clearly out of the allowed range\footnote{This statement only applies to off-diagonal Yukawa couplings. As explained at the beginning of this section, the $\mu_X$ limit, in which the $Y_\nu$ matrices are exactly diagonal, suppresses all cLFV observables and thus $\mathcal{O}(1)$ diagonal Yukawa couplings would be allowed.}.

\subsection{$\ell_i \to 3 \ell_j$}
\label{subsec:l3lresults}
 Concerning the 3-body decays  $\ell_i \to 3 \ell_j$ we focus on $\mu \to 3 e$,  due to its very challenging experimental bound. Since the $Z$-boson does not differentiate between leptonic flavours, the same behaviour and enhancement can be found for the cLFV $\tau$ 3-body decays into leptons.

\begin{figure}
\centering
\includegraphics[width=0.5\linewidth]{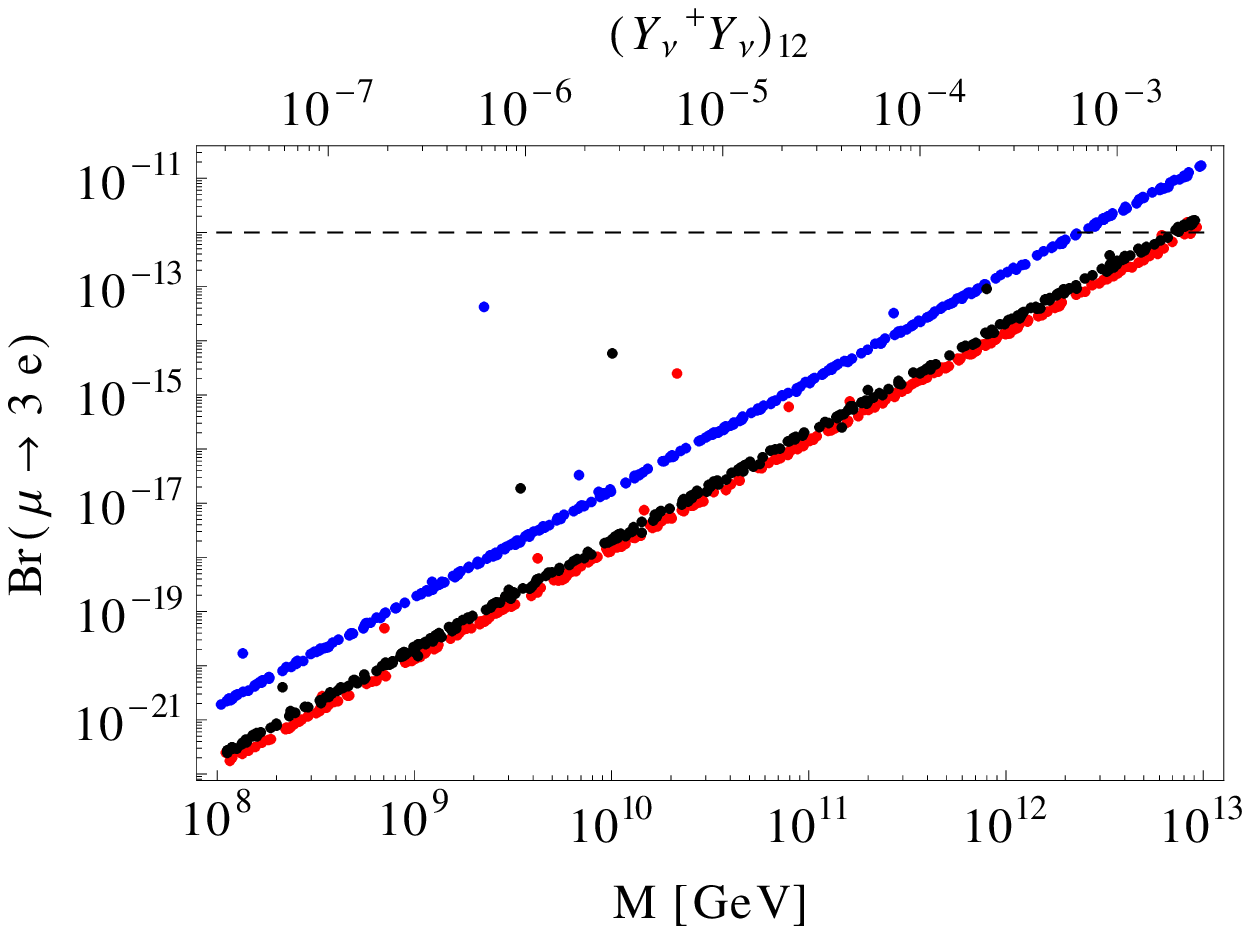}
\caption{$\text{Br}(\mu \to 3 e)$ as a function of $M$ and $(Y_\nu^\dagger Y_\nu)_{12}$ for different $\hat{M}_R$ values: $\hat{M}_R = 100$ GeV (blue), $\hat{M}_R = 1$ TeV (red) and $\hat{M}_R = 10$ TeV (black). The cMSSM parameters are taken as in figure \ref{CRfig}. The horizontal dashed line represents the current experimental bound.}
\label{BrMEEE-scatter}
\end{figure}

As discussed in \cite{Hirsch:2012ax}, the branching ratio for $\mu \to 3 e$ is greatly enhanced by the $Z$-penguin. In fact, it turns out that this process is only \emph{slightly} less constraining than $\mu-e$ conversion in nuclei. In Figure \ref{BrMEEE-scatter}, the $\text{Br}(\mu \to 3 e)$ is depicted as a function of $M$, and for three representative $\hat{M}_R$ values, again assuming  a degenerate singlet spectrum. 
As was the case of $\mu-e$ conversion in nuclei, the dependence 
on the cMSSM parameters 
is essentially negligible. Similarly, the scenario with $\hat{M}_R = 100$ GeV is dominated by the non-supersymmetric photonic penguins and larger branching ratios can thus be obtained. As can be seen,  the corresponding limits on $M$ and $(Y_\nu^\dagger Y_\nu)_{12}$ would be slightly  more relaxed than those arising from $\mu-e$ conversion in gold and titanium.

It is straightforward to obtain the following relation between the branching ratios of different $\ell_i \to 3 \ell_j$ channels
\begin{equation} \label{ratioBRs}
\frac{\Br(\ell_i \to 3 \ell_j)}{\Br(\ell_m \to 3 \ell_n)} = \frac{(Y_\nu^\dagger Y_\nu)_{ij}^2}{(Y_\nu^\dagger Y_\nu)_{mn}^2} \frac{m_{\ell_i}^5 \tau_i}{m_{\ell_m}^5 \tau_m}, 
\end{equation}
where $\tau_{i,m}$ are the life-times of the leptons. Equation \eqref{ratioBRs} provides a very good approximation to the numerical results and allows  to conclude that, unless strong cancellations occur in the $(Y_\nu^\dagger Y_\nu)_{ij}$ terms,  the three observables, $\Br(\mu \to 3 e)$, $\Br(\tau \to 3 e)$ and $\Br(\tau \to 3 \mu)$, are expected to lie within $1-2$ orders of magnitude. For example, assuming degenerate right-handed neutrinos, 
a vanishing Dirac phase in the neutrino sector, a normal hierarchy and a vanishing lightest neutrino mass ($m_{\nu_1} = 0$), one finds $\Br(\mu \to 3 e) \sim 70 \, \Br(\tau \to 3 e) \sim 0.7 \, \Br(\tau \to 3 \mu)$. An observation of $\tau \to 3 e$ or $\tau \to 3 \mu$ at rates much larger than those expected for $\mu \to 3 e$ could only be accommodated with a strong cancellation in $(Y_\nu^\dagger Y_\nu)_{12}$, that would suppress the $\mu-e$ transitions while still  allowing for $\tau-e$ and $\tau-\mu$ LFV\footnote{In such a scenario,  $\mu-e$ conversion in nuclei and $\mu \to e \gamma$ would be suppressed as well.}. Similar results  for this observable in an inverse seesaw framework with an extended gauge group have been found in  \cite{Hirsch:2012kv}. Finally, in $Z$-penguin dominated scenarios there is a clear correlation between the rates for $\mu \to 3 e$ and $\mu-e$ conversion in nuclei. Numerically, we find $\text{CR}(\mu-e,\text{Ti})/\Br(\mu \to 3e) \sim 15$.

\subsection{$\tau \to P^0 \ell_i$}
\label{subsec:taumesonlresults}
\begin{figure}
\centering
\includegraphics[width=0.5\linewidth]{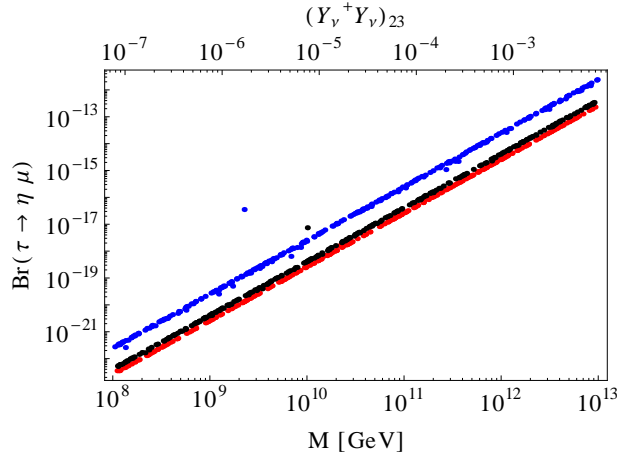}
\caption{$\text{Br}(\tau \to \eta \mu)$ as a function of $M$ and $(Y_\nu^\dagger Y_\nu)_{23}$ for different $\hat{M}_R$ values: $\hat{M}_R = 100$ GeV (blue), $\hat{M}_R = 1$ TeV (red) and $\hat{M}_R = 10$ TeV (black). The cMSSM parameters are taken as in figure \ref{CRfig}.}
\label{BrTauEtaMu-scatter}
\end{figure}

We now address $\tau$ flavour violating decays with a light pseudoscalar meson in the final state,  $\tau \to P^0 \ell_i$.
As done for the previous observables, we present our results for the branching ratios  as a function of $M$, and for three different  values for $\hat{M}_R$. This can be seen in Figure \ref{BrTauEtaMu-scatter}, where we focus on the particular case $\tau \to \eta \mu$.

We observe in Figure \ref{BrTauEtaMu-scatter} a clear enhancement which is again due to the dominance of the $Z$-boson  mediated contributions in the total amplitude.  However the branching ratio 
remains small as the size of the $(Y_\nu^\dagger Y_\nu)_{23}$ terms  never exceeds $O(10^{-2})$.

In fact, without the $Z$-penguin contributions, this observable would be much more suppressed (by several orders of magnitude). For example, the Higgs-mediated contributions  to this process were  studied in \cite{Abada:2011hm} and it was found that  $\mathcal{O}(1)$ Yukawa couplings would be  required to reach observable levels. However, the strong constraints set by $\mu-e$ conversion in nuclei preclude this possibility, implying that $\Br(\tau \to \eta \mu)\lesssim 10^{-13}$. In conclusion, if the inverse seesaw is realised in Nature, semileptonic $\tau$ cLFV decays do not have realistic chances of being observed in the near future unless strong cancellations occur in the $\mu-e$ sector.

Very similar results have been obtained for other mesons, as well as for final states including an electron instead of a muon. These observables also exhibit a very  little dependence on the spectrum: large variations in $m_0$ and $M_{1/2}$ have a minimal impact on the results for $\Br(\tau \to P^0 \ell_i)$.

\subsection{$\ell_i \to \ell_j \gamma$}
\label{subsec:megresults}

\begin{figure}
\centering
\includegraphics[width=0.5\linewidth]{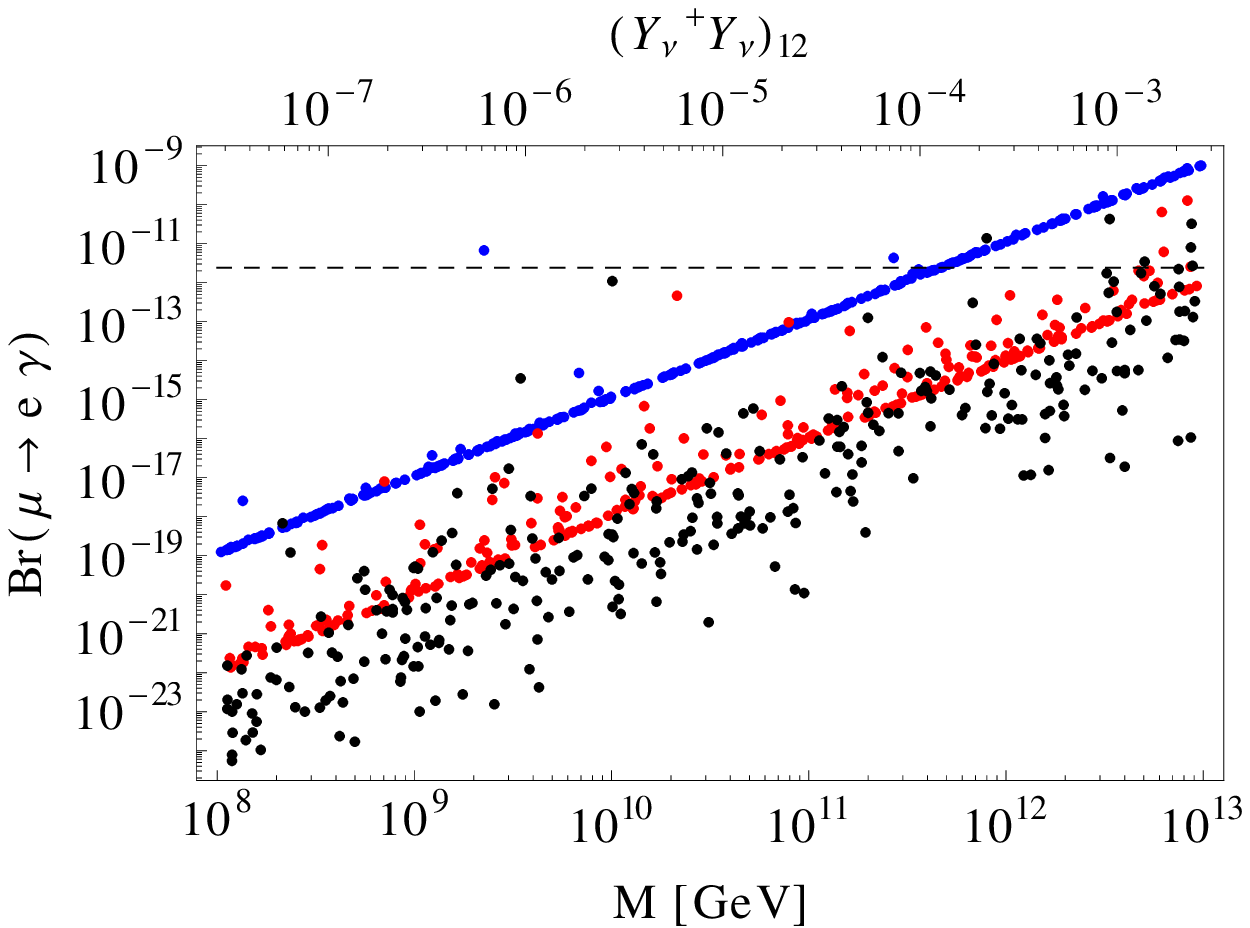}
\caption{$\Br(\mu \to e \gamma)$ as a function of $M$ and $(Y_\nu^\dagger Y_\nu)_{12}$ for different $\hat{M}_R$ values: $\hat{M}_R = 100$ GeV (blue), $\hat{M}_R = 1$ TeV (red) and $\hat{M}_R = 10$ TeV (black). The cMSSM parameters are taken as in figure \ref{CRfig}. The horizontal dashed line represents the current experimental bound.}
\label{BrMEG-scatter}
\end{figure}

Finally, we discuss our results for $\ell_i \to \ell_j \gamma$. Radiative decays have been intensively 
studied in the literature (see for instance~\cite{Hisano:1995cp,Hisano:1995nq,Deppisch:2004fa}). However, most of the phenomenological studies  have focused only on this observable, neglecting the impact of the $Z$-boson enhancement on the other cLFV observables. Therefore, our purpose is to determine how the constraints on the parameters of the inverse seesaw extension of the MSSM, derived from the analysis of the $Z$-boson enhanced observables, affect the predictions for the $\ell_i \to \ell_j \gamma$ rates.  
In fact, due to the strong limits on the Yukawa couplings coming from $\mu-e$ conversion in nuclei, $\Br(\ell_i \to \ell_j \gamma)$ is typically below the experimental limits,  the only exception being the $\hat{M}_R = 100$ GeV  scenario.  This can be  seen in Figure \ref{BrMEG-scatter} where, as expected from theoretical arguments,  $\text{Br}(\mu \to 3 e) > \text{Br}(\mu \to e \gamma)$ in the $Z$-penguin dominated scenarios\footnote{This is true for moderate values of $m_0$ and $M_{1/2}$ or, in other words, when the superparticles running in the loop are reasonably heavy. Similarly, low $M_R$ spoils this feature by increasing the non-SUSY contributions.}.

It is also clear from Figure \ref{BrMEG-scatter}  that for low $\hat{M}_R$, the observable $\Br(\mu \to e \gamma)$ has very little dependence on $m_0$ and $M_{1/2}$, whereas for large $\hat{M}_R$, one can find very large variations due to the different values of the SUSY masses. Similarly, we have checked that the other cMSSM parameters, $A_0$, $\tan \beta$, $\text{sign}(\mu)$ and $B_0$, do not  significantly affect the numerical value of $\Br(\mu \to e \gamma)$ when $\hat{M}_R = 100$ GeV,  but can become relevant for larger values of $\hat{M}_R$. 
We find that the non-SUSY contributions to $\Br(\mu \to e \gamma)$ become relevant only for $\hat{M}_R < 1$ TeV, and in fact, for $\hat{M}_R = 100$ GeV the non-SUSY contributions totally dominate, so that all dependence on $m_0$, $M_{1/2}$ and on the rest of cMSSM parameters disappears. These results are in agreement with those found in \cite{Deppisch:2004fa}, where it was shown that non-SUSY contributions can enhance $\Br(\ell_i \to \ell_j \gamma)$ in the inverse seesaw if the singlet fermions are light and have relatively large mixings with the active neutrinos. This result has been confirmed by our numerical study, where we found that $\Br(\ell_i \to \ell_j \gamma)$ could be enhanced by some orders of magnitude when $\hat{M}_R \ll m_W$. One can thus find regions of the $\hat{M}_R - M$ plane where $\text{Br}(\mu \to 3 e) < \text{Br}(\mu \to e \gamma)$, contrary to what occurs when SUSY contributions dominate in both 
observables\footnote{The $\hat{M}_R - M$ plane is not only constrained by cLFV observables, but also by Non-Standard Interactions (NSI) \cite{Antusch:2008tz}. 
In this model, for degenerate singlets, these bounds can be translated into $\mu_X \gtrsim 20$ eV. This in turn implies that, for a given value of $M$, one can always find a lower bound on $\hat{M}_R$. For example, for $M = 10^{12}$ GeV the NSI bound implies $\hat{M}_R \gtrsim 135$ GeV and thus the blue dots in Figure \ref{BrMEG-scatter} with $M \gtrsim 10^{12}$ GeV, which are ruled out by $\mu \to e \gamma$, are also disfavoured by NSI. For a non-degenerate spectrum the previous estimate does not hold.}.

\subsection{Other observables and benchmark points}
\label{res:summ}
To conclude the numerical analysis, we address some aspects not fully covered in the previous sections. We also summarise the most relevant features and results using  some 
 specific benchmark points.
\bigskip

Concerning flavour violating neutral boson decays, some additional
comments are in order.  The cLFV decays of the $Z$-boson, $Z \to
\ell_i \ell_j$ with $i \ne j$, are not enhanced\footnote{This process
  has been studied in detail in \cite{Bi:2000xp}, where the relevant
  expressions are given. Recently, good agreement, up to an overall
  factor of 2 in the wave function contributions, was found in
  \cite{Dreiner:2012mx}.} and the corresponding branching ratios lie
below experimental reach~\cite{Nakamura:2010zzi}.  Similarly, the
branching ratios for $H^0 \to \ell_i \ell_j$ are very small.  This
process is known to have a non-decoupling behaviour leading to
non-neglible contributions even for very heavy sparticles
\cite{Arganda:2004bz}. However, in the present case, the small Yukawa
couplings strongly suppress the corresponding rate. We have also
studied collider observables such as $\text{Br}(\tilde{\chi}_2^0 \to
\tilde{\chi}_1^0 \ell_i \ell_j)$ and slepton mass splittings, $\Delta
m_{\tilde{\ell}}$. We have found that they have very little relevance
in this model as expected from the previous discussion in Section
\ref{coll}.
\bigskip

\begin{figure}
\centering
\includegraphics[width=0.45\linewidth]{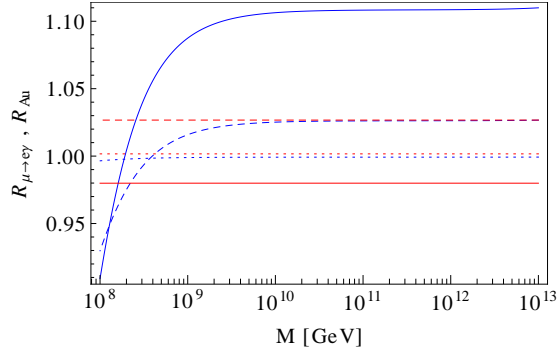}
\caption{$R_{\mu \to e \gamma}$ (blue) and $R_{\text{Au}}$ (red) as a function of $M$. The $M_R$ matrix is given by $M_R = \text{diag}(\hat{M}_{R 1},10 \, \text{TeV},10 \, \text{TeV})$, with $\hat{M}_{R 1} = 30$ GeV (solid lines), $\hat{M}_{R 1} = 100$ GeV (dashed lines) and $\hat{M}_{R 1} = 1$ TeV (dotted lines). The cMSSM parameters take the values $m_0 = 500$ GeV, $M_{1/2} = 1$ TeV, $A_0 = -300$ GeV, $B_0 = 0$ and $\tan \beta = 10$.}
\label{degcomp}
\end{figure}

For completeness, we briefly discuss the case of non-degenerate right-handed neutrinos, $M_R = \text{diag}(\hat{M}_{R 1},\hat{M}_{R 2},\hat{M}_{R 3})$, with  $\hat{M}_{R i} \ne \hat{M}_{R j}$. In our analysis we considered three different values for $\hat{M}_{R 1}$ ($30$ GeV, $100$ GeV and $1$ TeV), fixed $\hat{M}_{R 2} = \hat{M}_{R 3} = 10$ TeV and varied $M$ in the range [$10^8,10^{13}$ GeV]. In order to allow for a comparison  with the degenerate case (setting, for example, $\hat{M}_{R 1} = \hat{M}_{R 2} = \hat{M}_{R 3} = 10$ TeV) and to identify the effect of the right-handed neutrino spectrum on the cLFV observables, we have taken a common $M$ for the three right-handed neutrinos by adjusting the corresponding $\hat{\mu}_{X i}$. This allows to have identical values for the  Yukawa couplings in the degenerate and non-degenerate cases. To better  evaluate  the difference between the degenerate and non-degenerate cases, we further define the ratios
\begin{equation}\label{ratioMEG}
R_{\mu \to e \gamma} = \frac{\Br(\mu \to e \gamma)_{\text{non-deg}}}{\Br(\mu \to e \gamma)_{\text{deg}}} \quad , \quad R_{\text{Au}} = \frac{\text{CR}(\mu-e,{\text{Au}})_{\text{non-deg}}}{ \text{CR}(\mu-e,{\text{Au}})_{\text{deg}}}\ .
\end{equation}
 The results can be seen in Figure \ref{degcomp} where the cMSSM parameters take the arbitrary (but representative) values $m_0 = 500$ GeV, $M_{1/2} = 1$ TeV, $A_0 = -300$ GeV, $B_0 = 0$ and $\tan \beta = 10$. As expected, $\Br(\mu \to e \gamma)$ is affected by a change in the singlet spectrum. However, one can see that the relative difference (for this particular choice of parameters) can reach at most $\sim 10\%$. The change in the $Z$-boson mediated processes is negligible: for example, the relative change in the $\mu-e$ conversion rate in gold is always below $2\%$, independently of the value of $M$. This is another example of the little impact that the mass spectrum has on the $Z$-boson mediated processes. In conclusion, a non-degenerate right-handed neutrino spectrum can indeed 
induce changes to the observables, but these are typically small and the 
essence of the results derived here is unaffected by the nature of the 
right-handed neutrino spectrum\footnote{There is an exception to this general statement: in the non-degenerate case,  the $R$ matrix is indeed relevant even if
it is real \cite{Casas:2001sr}.}.

\bigskip

Finally, we present our results for the cLFV observables for the two
benchmark points of Table~\ref{tab:benchmark}. Point A is associated
to a Higgs mass $m_{h^0} = 126.5$ GeV, in agreement with the latest
LHC \cite{ATLAS:2012ae,Chatrchyan:2012tx} and Tevatron
\cite{TEVNPH:2012ab} results\footnote{Our numerical routines for the
  computation of the Higgs mass, based on {\tt SPheno 3.1.9}
  \cite{Porod:2003um,Porod:2011nf}, agree with the results of
  \cite{Arbey:2011ab,Baer:2011ab,Ellis:2012aa}, within the usual
  theoretical uncertainty of $\sim 1-2$ GeV.}. We note that, although
the inverse seesaw contains additional degrees of freedom beyond those
of the cMSSM, the smallness of the Yukawa couplings implies that any
correction to the Higgs mass, $m_{h^0}$, arising from the singlet
sector will be very small\footnote{For large Yukawa coupling, an
  enhancement to the Higgs mass can be found. However, this is often
  in conflict with bounds from LFV~\cite{Hirsch:2011hg}.}. Point B
exemplifies a very heavy supersymmetric spectrum with associated
gluino and squark masses larger than $6$ TeV and lightest chargino and
sneutrino masses of $2.3$ TeV and $3$ TeV, respectively. It has been
selected in order to show explicitly the non-decoupling behaviour of
the $Z$-mediated processes.

The results for the cLFV observables are given in Table~\ref{tab:benchmark-res} and,  for completeness, we display in Table \ref{tab:sensi} the current experimental
  bounds and future sensitivities for the same cLFV observables. For each point we have taken three different spectra in the singlet sector: (1) $\hat{M}_{R 1} = \hat{M}_{R 2} = \hat{M}_{R 3} = 100$ GeV, (2) $\hat{M}_{R 1} = \hat{M}_{R 2} = \hat{M}_{R 3} = 1$ TeV, and (3) $\hat{M}_{R 1} = 50$ GeV, $\hat{M}_{R 2} = 500$ GeV, $\hat{M}_{R 3} = 1$ TeV. The first case serves to show  the influence of the non-SUSY contributions to $\ell_i \to \ell_j \gamma$, the second one is a standard degenerate scenario where the SUSY contributions dominate all processes, while the third case is a non-degenerate scenario with an intermediate situation. In all cases we fix $M = 4 \times 10^{11}$ GeV.

 Firstly, we observe that $\Br(l_i \to \ell_j \gamma)$ shows very
 little change when going from scenario A-1 (with relatively light
 sparticles) to B-1 (with very heavy sparticles).  As discussed in
 Section \ref{subsec:megresults}, in scenarios with very low masses
 for the right-handed neutrinos, the main contributions to $\Br(l_i
 \to \ell_j \gamma)$ are non-supersymmetric. Therefore, they are
 obviously unaffected by the size of the SUSY parameters. We also
 point out that, although the right-handed neutrinos are light in
 these two scenarios, the bounds from NSI \cite{Antusch:2008tz} are
 satisfied due to the smallness of the Yukawa couplings. Another
 interesting remark that can be made in association to
 Table~\ref{tab:benchmark-res} is the different behaviour of the
 radiative $\ell_i \to \ell_j \gamma$ and the $Z$-mediated processes
 when going from the A-2 point to the B-2 point.  While $\Br(\ell_i
 \to \ell_j \gamma)$ is clearly reduced, in agreement with the
 well-known dependence on the SUSY spectrum, $\propto
 m_\text{SUSY}^{-4}$, the rates for the $Z$-mediated processes hardly
 change. This is a clear indication of the non-decoupling behaviour
 due to the $Z$-penguins. Finally, points A-3 and B-3 show an
 intermediate case, with one light right-handed neutrino, yielding
 results quite similar to those in points A-2 and B-2.  Thus, as
 discussed before, the nature of the right-handed neutrinos spectrum
 is irrelevant.

\begin{table}[tb!]
\centering
\begin{tabular}{|c|c c c c c c|}
\hline
Point & $m_0$ [TeV] & $M_{1/2}$ [TeV] & $A_0$ [TeV] & $B_0$ [TeV] & $\tan \beta$ & $\text{sign}(\mu)$ \\
\hline
\hline
A & 0.5 & 1.5 & -3 & 0 & 20 & + \\
\hline
B & 3 & 3 & 0 & 0 & 10 & + \\
\hline
\end{tabular}
\caption{Benchmark points used in the numerical analysis. Point A
  leads to a Higgs mass of $m(h^0) = 126.5$ GeV, in accordance with
  the latest LHC results. Point B is an example of a very heavy
  supersymmetric spectrum.}
\label{tab:benchmark} 
\end{table}

\begin{table}[tb!]
\centering
\begin{tabular}{|c||c|c|c||c|c|c|}
\hline
cLFV Process &  A-1 & A-2 & A-3 & B-1 & B-2 & B-3 \\
\hline
$\mu \to e \gamma$  & $1.9 \times 10^{-12}$ & $1.2 \times 10^{-13}$ & $1.3 \times 10^{-13}$ & $1.7 \times 10^{-12}$ & $1.5 \times 10^{-15}$ & $3.6 \times 10^{-15}$ \\
$\tau \to e \gamma$ & $2.9 \times 10^{-14}$ & $3.6 \times 10^{-15}$ & $3.4 \times 10^{-15}$ & $2.4 \times 10^{-14}$ & $2.1 \times 10^{-17}$ & $8.7 \times 10^{-17}$ \\
$\tau \to \mu \gamma$ & $2.7 \times 10^{-12}$ & $3.4 \times 10^{-13}$ & $3.3 \times 10^{-13}$ & $2.3 \times 10^{-12}$ & $2.0 \times 10^{-15}$ & $7.7 \times 10^{-16}$ \\
$\mu \to 3e$ & $2.9 \times 10^{-14}$ & $3.3 \times 10^{-15}$ & $2.7 \times 10^{-15}$ & $3.0 \times 10^{-14}$ & $2.4 \times 10^{-15}$ & $1.8 \times 10^{-15}$ \\
$\tau \to 3e$ & $5.5 \times 10^{-16}$ & $7.6 \times 10^{-17}$ & $1.3 \times 10^{-16}$ & $5.3 \times 10^{-16}$ & $3.4 \times 10^{-17}$ & $9.2 \times 10^{-17}$ \\
$\tau \to 3 \mu$ & $2.9 \times 10^{-14}$ & $4.4 \times 10^{-15}$ & $4.7 \times 10^{-15}$ & $3.1 \times 10^{-14}$ & $3.2 \times 10^{-15}$ & $3.6 \times 10^{-15}$ \\
$\mu-e$ , Au & $1.2 \times 10^{-13}$ & $3.5 \times 10^{-14}$ & $2.7 \times 10^{-14}$ & $1.4 \times 10^{-13}$ & $2.8 \times 10^{-14}$ & $2.1 \times 10^{-14}$ \\
$\mu-e$ , Ti & $6.7 \times 10^{-14}$ & $2.8 \times 10^{-14}$ & $2.2 \times 10^{-14}$ & $8.4 \times 10^{-14}$ & $2.2 \times 10^{-14}$ & $1.6 \times 10^{-14}$ \\
$\tau \to e \eta$ & $4.3 \times 10^{-17}$ & $4.5 \times 10^{-18}$ & $1.3 \times 10^{-17}$ & $4.6 \times 10^{-17}$ & $4.7 \times 10^{-18}$ & $1.3 \times 10^{-17}$ \\
$\tau \to \mu \eta$ & $4.0 \times 10^{-15}$ & $4.2 \times 10^{-16}$ & $4.7 \times 10^{-16}$ & $4.3 \times 10^{-15}$ & $4.3 \times 10^{-16}$ & $4.9 \times 10^{-16}$ \\
\hline
\end{tabular}
\caption{Results for several lepton flavour violating processes for
  the benchmark points of Table~\ref{tab:benchmark} supplemented with
  three different singlet spectra: (1) $\hat{M}_{R 1} = \hat{M}_{R 2}
  = \hat{M}_{R 3} = 100$ GeV, (2) $\hat{M}_{R 1} = \hat{M}_{R 2} =
  \hat{M}_{R 3} = 1$ TeV, and (3) $\hat{M}_{R 1} = 50$ GeV,
  $\hat{M}_{R 2} = 500$ GeV, $\hat{M}_{R 3} = 1$ TeV.}
\label{tab:benchmark-res}
\end{table}

\begin{table}[tb!]
\centering
\begin{tabular}{|c|c|c|}
\hline
cLFV Process & Present Bound & Future Sensitivity  \\
\hline
$\mu \to e \gamma$ & $2.4 \times 10^{-12}$ \cite{PhysRevLett.107.171801} & $\mathcal{O}(10^{-13})$ \cite{PhysRevLett.107.171801}  \\
$\tau \to e \gamma$ & $3.3 \times 10^{-8}$ \cite{PhysRevLett.104.021802}& $3.0 \times 10^{-9}$ \cite{OLeary2010af}\\
$\tau \to \mu \gamma$ & $4.4 \times 10^{-8}$ \cite{PhysRevLett.104.021802}& $2.4 \times 10^{-9}$ \cite{OLeary2010af} \\
$\mu \to 3 e$ & $1.0 \times 10^{-12}$\cite{Bellgardt:1987du} & $\mathcal{O}(10^{-16})$ \cite{LOImu3e}\\
$\tau \to 3 e$ & $2.7\times10^{-8}$\cite{Hayasaka:2010np} & $2.3 \times 10^{-10}$ \cite{OLeary2010af}  \\
$\tau \to 3 \mu$ & $2.1\times10^{-8}$\cite{Hayasaka:2010np} & $8.2 \times 10^{-10}$ \cite{OLeary2010af}  \\
$\mu-e$ , Au & $7.0 \times 10^{-13}$ \cite{Bertl:2006} & $ $  \\
$\mu-e$ , Ti & $4.3 \times 10^{-12}$ \cite{Dohmen:1993mp} & $\mathcal{O}(10^{-18})$ \cite{PRIME} \\
$\tau \to e \eta$ & $4.4\times 10^{-8}$\cite{arXiv:1011.6474} & $\mathcal{O}(10^{-10})$ \cite{OLeary2010af}\\
$\tau \to \mu \eta$ & $2.3\times 10^{-8}$\cite{arXiv:1011.6474} & $\mathcal{O}(10^{-10})$ \cite{OLeary2010af}\\
\hline
\end{tabular}
\caption{The current experimental
  bounds and future sensitivities for the cLFV observables.}
\label{tab:sensi}
\end{table}

\section{Conclusion}
\label{concs}
The supersymmetric inverse seesaw is a very attractive extension of
the MSSM, with neutrino masses generated by TeV-scale mediators, in
association with potentially large Yukawa couplings. In this work, we
have studied in detail the predictions for several lepton flavour
violating observables, focusing on those mediated by $Z$- boson
exchange. These are particularly interesting when the
non-supersymmetric contributions are small.

We have found that, due to the  non-decoupling behaviour, the
$Z$-penguins totally dominate the cLFV amplitudes in most  of
the parameter space, especially in scenarios where the right-handed
neutrinos and the supersymmetric particles have masses larger than
$\sim 500$ GeV. In those cases, $\mu-e$ conversion in nuclei is the
most constraining observable, clearly more restrictive than $\mu \to e
\gamma$. As a result of the large enhancement provided by the
$Z$-penguins, one can set very strong constraints on the size of the
flavour violating couplings. 

Finally, we emphasize that the bounds obtained in this work apply to
the off-diagonal elements of the Yukawa couplings. One can also set
bounds on the diagonal ones using the invisible width of the $Z$
boson.  This effect is negligible in our case due to the smallness of
the Yukawa couplings.

\subsection*{Acknowledgements}
We are grateful to Florian Staub for comments and corrections on the
numerical code. A.V. thanks Mar\'ia Jos\'e Herrero for interesting
discussions. D.D. acknowledges financial support from the CNRS. This
work has been partly done under the ANR project CPV-LFV-LHC
NT09-508531.  The authors acknowledge partial support from the
European Union FP7 ITN INVISIBLES (Marie Curie Actions, PITN- GA-2011-
289442).

\end{document}